\pdfoutput=1

\documentclass[pdflatex,sn-nature]{sn-jnl}

\usepackage{graphicx}%
\usepackage{multirow}%
\usepackage{amsmath,amssymb,amsfonts}%
\usepackage{xcolor}%
\usepackage{textcomp}%
\usepackage{eurosym}%
\usepackage{adjustbox}%
\usepackage{booktabs}%
\usepackage{array}%
\usepackage{makecell}%
\usepackage{longtable}%
\usepackage{algorithm}%
\usepackage{algorithmicx}%
\usepackage{algpseudocode}%
\usepackage[title]{appendix}%
\usepackage{tikz}%
\usetikzlibrary{positioning,arrows.meta,fit,backgrounds,calc,shapes.geometric,shapes.misc,decorations.pathreplacing}%
\usepackage{pgfplots}%
\pgfplotsset{compat=1.18}%
\usepgfplotslibrary{groupplots}%
\usepackage{hyperref}%

\graphicspath{{./}}

\definecolor{scA}{HTML}{8FB0CE}
\definecolor{scB}{HTML}{2F5C8A}
\definecolor{scC}{HTML}{1F7A74}
\definecolor{scD}{HTML}{B7791F}
\definecolor{scE}{HTML}{8C3B2F}
\definecolor{scStl}{HTML}{2F5C8A}
\definecolor{scTl}{HTML}{1F7A74}
\definecolor{scAm}{HTML}{B7791F}
\definecolor{scLt}{HTML}{5FA8A0}
\definecolor{scRd}{HTML}{8C3B2F}
\definecolor{scRust}{HTML}{A05A2C}
\definecolor{covfill}{HTML}{2F5C8A}
\definecolor{covempty}{HTML}{E3E7EC}

\definecolor{inkblue}{HTML}{1F3A5F}
\definecolor{steel}{HTML}{2F5C8A}
\definecolor{teal}{HTML}{1F7A74}
\definecolor{amber}{HTML}{D9881F}
\definecolor{slate}{HTML}{38414F}
\definecolor{paleblue}{HTML}{EAF1F8}
\definecolor{paleteal}{HTML}{E2F2F0}
\definecolor{paleamber}{HTML}{FBF1DC}
\definecolor{palegray}{HTML}{EEF0F2}
\definecolor{linegray}{HTML}{B7BFC9}
\definecolor{amberfill}{HTML}{F6D6A0}

\newlength{\cellw}
\newlength{\cellh}
\newlength{\cpad}
\newlength{\gut}
\newlength{\descw}
\newlength{\ewidth}
\newlength{\rowoff}

\newcommand{\cellbody}[5]{%
  \node[badge] (#1-b) at ([shift={(\cpad,-\cpad)}]#1.north west) {#2};
  \node[celltitle, anchor=west] at ([xshift=6pt]#1-b.east) {#3};
  \node[celldesc, anchor=north west, text width=#5]
       at ([shift={(\cpad,-\rowoff)}]#1.north west) {#4};
}

\begin{document}

\title[Agentic orchestration for rapid multi-scenario crisis analysis]{Agentic AI orchestration of heterogeneous economic models for rapid, multi-scenario analysis of energy crises}

\author*[1]{\fnm{Dana} \sur{Golden}}\email{dana.golden@stonybrook.edu}
\author[2]{\fnm{Brett} \sur{Indelicato}}\email{brett.indelicato@stonybrook.edu}
\author[3]{\fnm{Lav R.} \sur{Varshney}}\email{lav.varshney@stonybrook.edu}
\author[4]{\fnm{Carlos D.} \sur{Messina}}\email{cmessina@ufl.edu}
\author[4]{\fnm{Suzanne} \sur{Thornsbury}}\email{thornsbs@ufl.edu}

\affil*[1]{\orgdiv{Center of Excellence in Wireless and Information Technology (CEWIT)}, \orgname{Stony Brook University}, \orgaddress{\city{Stony Brook}, \state{New York}, \country{United States}}}
\affil*[2]{\orgdiv{Department of Economics}, \orgname{Stony Brook University}, \orgaddress{\city{Stony Brook}, \state{New York}, \country{United States}}}
\affil[3]{\orgdiv{AI Innovation Institute}, \orgname{Stony Brook University}, \orgaddress{\city{Stony Brook}, \state{New York}, \country{United States}}}
\affil[4]{\orgdiv{Institute for Food and Agricultural Studies}, \orgname{University of Florida}, \orgaddress{\city{Gainesville}, \state{Florida}, \country{United States}}}

\abstract{Rigorous economic models can take months to construct, yet energy crises demand decisions from policymakers within days or even hours. Any disruption in energy markets is not isolated but rapidly disseminates through interlinked global systems. Off-the-shelf models that already exist typically focus only on limited aspects of the system and are distributed across research groups, programming languages, software architectures not designed for model integration, and incompatible formats. Integrating these models manually can take longer than the crisis itself, forcing analysts to rely on whichever models are easiest to connect and leaving consequential scenarios unexplored. Policymakers must make rapid decisions with obstructed and limited information.

We show that large language models can perform the critical integration directly, revealing connections in a way that is timely and computationally efficient. The system constructs internally consistent scenarios, translates assumptions into model-specific inputs, executes existing economic and physical models in dependency order, and synthesizes outputs tailored to policymakers. The language model generates no quantitative results: every reported value is reproduced directly from an underlying model run, remains traceable to its source and is subject to analyst approval at each stage.

We develop a LLM framework that coordinates 16 models of oil, natural gas, shipping, water, helium, fertilizer and macroeconomic equilibrium. The framework is applied across five scenarios to assess the 2026 closure of the Strait of Hormuz and refreshed weekly for eight weeks as events on the ground continued to unfold.
Results revealed effects that isolated analyses would miss: the largest oil-price increase did not occur in the scenario with the greatest overall economic damage, and escalation generated a potential humanitarian threat to Gulf desalination systems. By linking models that already exist and reading them as a suite rather than in isolation, this architecture mobilizes distributed scientific models rapidly during energy and geopolitical disruptions while keeping any single model's assumptions from driving the conclusion.
}


\maketitle

\section{Introduction}\label{sec:intro}

Energy-supply shocks create a fast-moving crisis that literally and figuratively disrupts systems of power. Consequences can be extreme in complexity, reach, and destabilizing effect. The oil crises of 1973 and 1979 and the European gas crisis of 2022 elevated energy supply decisions into inflation, food security, and geopolitics within weeks, faster than conventional modelling cycles could follow.

When the Strait of Hormuz closed in 2026, the pattern repeated: policymakers had to respond in real time as a disruption to the energy sector rippled throughout multiple components of the global economy. The response required was not one-and-done: the crisis demanded a sustained analytical effort, which is why our analysis continued over eight weeks as events unfolded.

The crisis highlights a paradox: the analytical models needed to support policymaker decisions usually already exist; however, they were developed to focus on single parts of the problem and are slow to adapt and deploy. For example, the world oil-market model exists in one research group's code. The computable general equilibrium (CGE) model of global trade lives in another's. The regional hydrological model, the shipping-network model, and the energy-systems optimizer each live somewhere else entirely, written in different languages, requiring different inputs, and emitting incompatible outputs. Existing models are built slowly and carefully to answer big questions after deep exploration, but the most important questions in crises require speed, interoperability of models, and adaptability from analysis. Moreover, the global economy is a complex system whose aggregate behavior is hard to predict from its parts alone; it requires the models of those parts to interact so that emergent behaviors and outcomes can be captured.

Crisis analysis therefore fails not due to a lack of models but for lack of time to connect them and to let them interact to create a picture of the whole. An analyst supporting decision making in real time is confined to the frameworks at hand, forced to narrow the question to what those frameworks can answer, and limited to a handful of hand-built scenarios. The inflection points and cross-domain interactions that matter most are precisely what this narrowing discards.

A structural divide in economic modelling compounds the problem. Short-run partial-equilibrium commodity models capture the substitution possibilities and institutional detail that macroeconomic models abstract away; however, only a general-equilibrium treatment captures the interactions among energy prices, inflation, monetary policy, and investment that determine aggregate outcomes. The economic and strategic consequences of a conflict are modelled through approaches that rarely meet \cite{bib_hourcade2006}.

Rapid crisis analysis is not hindered by a shortage of models but by the lack of orchestration: the manual labor of making heterogeneous models interoperate, of swapping one model or scenario for another as the question shifts, and of chaining one model's output into another's input across scales. Because that labor grows with the number of models and scenarios, it forces the narrowing that conventional crisis analysis exhibits.

If the marginal cost of orchestration can be reduced, the speed and breadth of analysis that was previously infeasible becomes routine. Decision-makers can apply existing modelling capabilities in real time across many scenarios and points of critical uncertainty. If policymakers want to game out escalation, they can rerun the analysis cheaply; when the situation evolves, they can adapt the scenarios to match. The ability to integrate models helps manage a complex system whose long-term behavior is difficult to predict.

Analytical capability built over months can be repurposed within hours to move in lockstep with the crisis. Two payoffs follow. Speed moves rigorous analysis close enough to the decision to inform it, rather than arriving after the choice is made. Breadth replaces point forecasting with a map of possible outcomes across many scenarios, which raises the odds of catching the unintended consequences that any single scenario would conceal.

We present an agentic framework designed around exactly these properties. By agentic we mean something narrower than autonomy: a large language model (LLM) orchestrates a structured workflow rather than generating answers. Every domain model is wrapped once in a thin, standardized layer and entered in a registry. The LLM composes internally consistent scenario narratives, extracts quantitative assumptions, dispatches the models level by level, and synthesizes their outputs. Adding, removing, or substituting a model means changing a registry entry, not re-engineering a pipeline. A low marginal cost of substitution is the architecture's defining property.

Crisis analysis is among the most sensitive of domains: the cost of being wrong can be enormous. To ensure system outputs are valid, two commitments discipline the automation. First, analysts sign off at three mandatory checkpoints, and the pipeline is never run unattended. A human is in the loop to verify at every stage. Second, a non-fabrication rule confines the LLM to prose: policymakers can choose and edit specific scenario parameters of interest, and every quantitative value is passed through unmodified from a model run, with a provenance chain recording its path back through model output, model input, extracted parameter, and source narrative. The division of labor is strict. The models produce the numbers; the LLM produces the connective tissue.

A second contribution follows from running many models rather than one. Every model embeds assumptions about substitution, market structure, and behavior that suit the question it was built to answer. In a single-model study, assumptions pass silently into the results. A portfolio does not remove them, but it keeps any one model's assumptions from dictating full outcomes and turns disagreements between models into information for decision-makers. Because each number carries its provenance, the analyst can trace a divergence or a surprising result back to the model and the assumption that produced it, rather than averaging the difference away.

We demonstrate our framework on the 2026 closure of the Strait of Hormuz, coordinating sixteen heterogeneous models across five scenarios and re-deriving the full analysis weekly as the crisis evolved. The result is not a table of point estimates but an auditable reading of the crisis by region, over time, and across scenarios. Throughout, the crisis is the proving ground and the orchestration capability is the object of study. The architecture is designed to transfer to a wide class of energy and geopolitical disruptions.

\section{A framework built for substitution}\label{sec:framework}

Three operations dominate the labor of a typical multi-model analysis: interoperating across heterogeneous models, chaining outputs across scales, and substituting models and scenarios. We focus on gaining efficiencies in substitution in order to convert a one-off analysis into a living one. Every analysis becomes a reproducible framework rather than an ad hoc exercise.

\subsection{Three operations, one interface}\label{subsec:three_ops}

For interoperability, an LLM agent reads, writes, and translates between the input and output formats of models written in Python, Julia, R, GAMS, and spreadsheets. In our implementation, each model is wrapped exactly once, and the wrapper's declared input and output specification enforces interoperability rather than an ad hoc pairing.

Two design choices make substitution cheap. Introducing a domain model means writing one wrapper against a fixed interface and registering it; nothing else in the orchestration layer changes. Re-specifying a scenario, retuning a cross-model linkage, or adjusting a consistency tolerance can be achieved by changing a single setting. Because substitution is inexpensive, the analyst can fold in new information as a crisis unfolds and can evaluate far more models and scenarios than traditional approaches would permit. Using a diverse set of models, or combining them and varying their initial parameterizations across the range of estimates reported in the literature, lets analysts build ensembles that better characterize the consequential paths a crisis may follow.

The agent then chains runs across scales; for example, forwarding combat-model outputs into commodity models and commodity outputs into macroeconomic models. In this way, we provide a micro-to-macro analysis that remains both internally consistent and faithful to the underlying dynamics of the constituent models. Where one model has already computed a quantity that is input to another model, the framework forwards that value rather than asking the LLM to re-derive it. Prior inputs are inspectable and auditable; the non-fabrication rule and the three analyst checkpoints introduced above apply at every stage.

The framework does not replace the analyst. It empowers the analyst by removing the operational burden of running models and moving data between them, and by providing access to newly integrated systems and ensembles. This allows human judgment to concentrate where human analysts have a comparative advantage: deciding which assumptions are credible, which scenarios are plausible, which outputs are trustworthy, and how results should inform a decision. Analysts can focus on reading news, performing deep research, and talking to sources. This division of labor assigns to the machine the tasks that are operationally intensive but conceptually routine, consistent with the principle that automation is best deployed on work that is dull, dangerous, and dirty \cite{bib_dod_ddd}. Rather than replacing the critical functions of analysts, the system lets them concentrate on the most interesting and useful parts of the analysis.

\subsection{Scenario architecture}\label{subsec:scenario_arch}

Scenario construction follows a structured methodology in the tradition of Schwartz \cite{bib_schwartz}. The analyst fixes a focal issue, enumerates the key factors and driving forces that shape it, separates predetermined elements that appear in every scenario from the critical uncertainties that distinguish them, and crosses the two most consequential uncertainties into a matrix of internally consistent scenario worlds. A prescribed scenario can be generated outside the matrix when a decisive contingency is identified outside the matrix.

For each scenario the LLM generates a narrative timeline, extracts quantitative parameters with analyst input and verification, and maps them to each model's input specification. After execution, the agent collects outputs, checks internal consistency, and produces a structured synthesis for analyst review.

\section{The 2026 closure of the Strait of Hormuz}\label{sec:demo}
The 2026 closure of the Strait of Hormuz presents a demanding test case for the framework. Few disruptions have evolved as rapidly or reached as many commodity markets at once. The Strait is the most strategically significant maritime chokepoint for energy trade and carries substantial trade in other key segments. Before the 2026 closure, the narrow passage between Iran and Oman carried roughly one-fifth of global oil consumption and a comparable share of traded liquefied natural gas (LNG) \cite{bib_eia_hormuz}. Because Gulf export infrastructure is concentrated across Saudi Arabia, Iraq, Kuwait, Qatar, and the UAE, disruptions are global.

Tensions that escalated through 2025 culminated on 28 February 2026 in a coordinated air campaign by U.S. and Israeli forces against Iranian targets, including sites along the Strait \cite{bib_hormuz_strikes,bib_eia_q1_2026}. Shortly afterward the Islamic Revolutionary Guard Corps warned commercial vessels that the Strait was no longer secure. Regional shipping traffic fell by roughly 70\% within hours. Direct attacks on merchant vessels over the following days confirmed an operational closure, formally acknowledged on 2 March 2026 \cite{bib_hormuz_strikes}.

The market consequences were immediate. About 20 million barrels per day of oil flow and roughly 22\% of global LNG trade were frozen, and major Gulf producers including Saudi Arabia, Iraq, and the UAE shut in output \cite{bib_reuters_shutins}. Brent crude rose 10--13\% to above \$80 per barrel within days, surpassed \$100 on 12 March, and closed the first quarter near \$118 as front-month futures \cite{bib_eia_q1_2026}. Physical Brent cargoes traded higher still, and front-month Brent later touched roughly \$126 intraday on 30 April 2026, its highest level since April 2022 \cite{bib_brent_peak}. In Europe the shock met depleted winter storage, and Dutch Title Transfer Facility (TTF) gas benchmarks nearly doubled toward \texteuro{}60/MWh \cite{bib_ttf_gas}. Rerouting around the Cape of Good Hope added roughly 10--15 days to Asia--Europe transit, pushed projected freight rates up 25--30\%, and sharply raised war-risk insurance, which in the first days after closure roughly doubled to about \$250{,}000 per supertanker voyage \cite{bib_warrisk}.

\subsection{Five commodity systems}\label{subsec:five_systems}

The crisis runs through five primary commodity systems with distinct downstream paths: oil, LNG, fertilizer, helium, and water. The United States sits on the supply side of the first four as the largest oil and gas producer, a major LNG and fertilizer exporter, and the holder of the world's largest helium reserves. Water follows a different path as a localized, life-sustaining commodity that is often treated as a public utility. Gulf states depend on desalination for potable water, so damage to those facilities carries a humanitarian risk that we argue has often been underpriced by market analysis throughout the conflict.

The outcomes tracked in our demonstration are organized along two dimensions, time horizon and scope, so that short-run price adjustment and medium-to-long-run structural change are followed together (Table~\ref{tab:outcomes}). This outcome set is specific to the demonstration; the framework itself is agnostic to it.

\begin{table}[t]
\caption{Outcomes of interest by time horizon and scope.}\label{tab:outcomes}
\begin{tabular*}{\textwidth}{@{\extracolsep\fill}lll}
\toprule
Scope & Short-run & Long-run \\
\midrule
Micro & Oil, LNG, fertilizer, helium prices & Regional water sustainability \\
 & Crop yields and prices & LNG and electricity market health \\
 & Desalination capacity adequacy & Semiconductor industry growth \\
\midrule
Macro & Inflation; GDP growth & Fiscal sustainability \\
 & Interest rates and debt service & Unemployment; trade deficit \\
 & AI-related investment & Recession probability \\
\midrule
Strategic & Combat objective attainment & Reserve-currency status \\
 & Alliance maintenance & Weapons-stockpile adequacy \\
 & Critical-infrastructure protection & Industrial-base health \\
\botrule
\end{tabular*}
\end{table}

\subsection{Scenarios and the analytical-level stack}\label{subsec:scenarios}

\begin{figure*}[t]
\centering
\begin{adjustbox}{max width=\textwidth}
\begin{tikzpicture}[
    font=\sffamily,
    cell/.style     ={draw=linegray, line width=0.7pt, rounded corners=3pt,
                      minimum width=\cellw, minimum height=\cellh, anchor=south west},
    badge/.style    ={circle, fill=slate, text=white, font=\sffamily\bfseries,
                      minimum size=19pt, inner sep=0pt},
    celltitle/.style={font=\sffamily\bfseries\small, text=slate, inner sep=0pt},
    celldesc/.style ={font=\sffamily\footnotesize, text=slate, align=left, inner sep=0pt},
    axislab/.style  ={text=slate, font=\sffamily\bfseries\small},
  ]
  \node[cell, fill=paleblue]  (A) at (0,0)                       {};
  \node[cell, fill=paleamber] (C) at (\cellw+\gut,0)             {};
  \node[cell, fill=paleteal]  (B) at (0,\cellh+\gut)             {};
  \node[cell, fill=amberfill] (D) at (\cellw+\gut,\cellh+\gut)   {};
  \cellbody{A}{A}{Swift\,/\,Contained}%
    {Sharp but transient shock; markets clear within weeks.}{\descw}
  \cellbody{B}{B}{Prolonged\,/\,Contained}%
    {Rerouting, reserve draws, demand destruction; large adjustment costs.}{\descw}
  \cellbody{C}{C}{Swift\,/\,Escalated}%
    {Strait reopens, but conflict widens; severe infrastructure and humanitarian costs.}{\descw}
  \cellbody{D}{D}{Prolonged\,/\,Escalated}%
    {Sustained closure plus regional conflict; outer bound of the matrix.}{\descw}
  \draw[-{Stealth[length=2.6mm]}, line width=0.9pt, slate]
        ([yshift=-14pt]A.south west) -- ([yshift=-14pt]C.south east);
  \node[axislab] at ($(A.south)!0.5!(C.south)+(0,-30pt)$)
        {Scope of conflict escalation};
  \draw[-{Stealth[length=2.6mm]}, line width=0.9pt, slate]
        ([xshift=-14pt]A.south west) -- ([xshift=-14pt]B.north west);
  \node[axislab, rotate=90] at ($(A.west)!0.5!(B.west)+(-32pt,0)$)
        {Duration of Strait closure};
  \coordinate (Emid) at ($(C.north east)!0.5!(D.south east)$);
  \node[draw=amber, line width=1pt, dashed, rounded corners=3pt, fill=white,
        anchor=west, minimum width=\dimexpr\ewidth+2\cpad\relax, minimum height=3.05cm]
        (E) at ([xshift=26pt]Emid) {};
  \cellbody{E}{E}{Infrastructure collapse}%
    {\emph{Prescribed tail risk, outside the $2{\times}2$ matrix.} GCC desalination,
     the Qatar LNG complex, and major Gulf oil terminals are destroyed, with a
     multi-year reconstruction overhang.}{\ewidth}
  \draw[-{Stealth[length=2.2mm]}, amber, dashed, line width=1pt]
        ([xshift=2pt]Emid) -- ([xshift=24pt]Emid);
\end{tikzpicture}
\end{adjustbox}
\caption{Scenario space. Two critical uncertainties, the duration of the Strait closure and the scope of conflict escalation, define a two-by-two matrix of internally consistent scenarios (A to D). A fifth tail-risk scenario (E), an infrastructure collapse affecting Gulf desalination, the Qatar LNG complex, and major oil terminals, is prescribed outside the matrix.}
\label{fig:scenario_space}
\end{figure*}
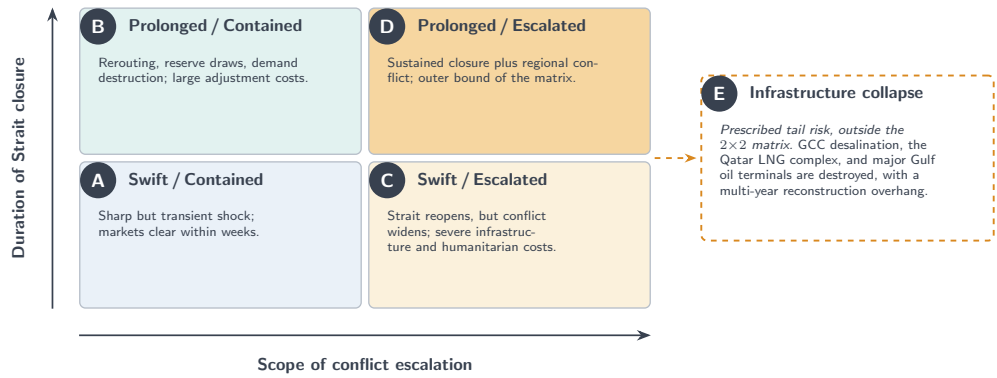

We structure the scenario space with two critical uncertainties: the duration of the closure and the scope of conflict escalation. Crossing them yields four matrix scenarios (A to D), to which we add a fifth, off-matrix tail-risk scenario (E) that stresses the framework's handling of regional infrastructure destruction (Fig.~\ref{fig:scenario_space}).

Scenario A is a swift, contained closure; B a prolonged but contained closure; C a swift reopening amid wider escalation; and D a prolonged closure with broad escalation. Scenario E is an infrastructure-collapse tail in which Gulf desalination, the Qatar LNG complex, and major oil terminals are destroyed with a multi-year reconstruction overhang. Scenario E is authored rather than drawn from the matrix, because its premises are predetermined elements the matrix is not designed to vary.

\begin{figure*}[t]
\centering
\resizebox{\textwidth}{!}{%
\begin{tikzpicture}[font=\sffamily]
  \tikzset{
    band/.style={draw=linegray, line width=0.7pt, rounded corners=3pt,
      minimum width=12cm, align=left, anchor=center},
    chip/.style={draw=linegray, fill=white, rounded corners=1.5pt, line width=0.5pt,
      font=\sffamily\scriptsize, align=center, inner sep=3pt, minimum height=0.92cm},
    lvl/.style={font=\sffamily\bfseries\small, align=right, anchor=east, text width=2.7cm},
    fwd/.style={font=\sffamily\footnotesize\itshape, slate, align=center,
      fill=white, inner sep=2pt},
    cluster/.style={font=\sffamily\scriptsize\itshape, slate, anchor=center},
    flow/.style={-{Stealth[length=2.6mm]}, line width=1pt, steel},
  }
  \def\bx{1.6}
  \def\ycombat{10.2}
  \def\ycom{7.825}
  \def\ydown{5.45}
  \def\ysrm{3.45}
  \def\ylrm{0.95}
  \node[band, fill=palegray, minimum height=1.0cm]  (combat) at (\bx,\ycombat){};
  \node[band, fill=paleblue, minimum height=1.75cm] (com)    at (\bx,\ycom){};
  \node[band, fill=paleteal, minimum height=1.0cm]  (down)   at (\bx,\ydown){};
  \node[band, fill=paleamber, minimum height=1.0cm] (srm)    at (\bx,\ysrm){};
  \node[band, fill=amber!16, minimum height=2.0cm]  (lrm)    at (\bx,\ylrm){};
  \node[lvl] at (combat.west|-combat) {Combat\\\footnotesize\mdseries physical events};
  \node[lvl] at (com.west|-com)       {Commodity\\\footnotesize\mdseries markets};
  \node[lvl] at (down.west|-down)      {Commodity-\\downstream};
  \node[lvl] at (srm.west|-srm)        {Short-run\\macro};
  \node[lvl] at (lrm.west|-lrm)        {Long-run macro\\\& energy systems};
  \node[chip, text width=4.0cm] at (\bx,\ycombat)
        {infrastructure damage,\\capability degradation};
  \node[chip, text width=2.0cm] at (-3.00,\ycom) {\textbf{Oil}\\BKR\,$\cdot$\,POLES\\MarketSim};
  \node[chip, text width=2.0cm] at (-0.70,\ycom) {\textbf{LNG}\\GGM\,$\cdot$\,LNGST\\Energy Flux};
  \node[chip, text width=2.0cm] at ( 1.60,\ycom) {\textbf{Helium}\\World\\Helium};
  \node[chip, text width=2.0cm] at ( 3.90,\ycom) {\textbf{Fertilizer}\\World Fert.\\futures};
  \node[chip, text width=2.0cm] at ( 6.20,\ycom) {\textbf{Water}\\CWatM\,$\cdot$\,WEAP\\SahysMod};
  \node[chip, text width=2.6cm] at (-0.20,\ydown) {SimRLFab (fab throughput)};
  \node[chip, text width=2.9cm] at ( 3.25,\ydown) {Argonne Helium ABM};
  \node[chip, text width=1.6cm] at (-0.50,\ysrm) {NEMS};
  \node[chip, text width=1.6cm] at ( 1.60,\ysrm) {MAM};
  \node[chip, text width=1.6cm] at ( 3.70,\ysrm) {NREL};
  \node[cluster] at ( 0.05,1.55) {general equilibrium};
  \node[cluster] at ( 5.84,1.55) {energy systems (long-run)};
  \node[chip, text width=1.7cm] at (-3.20,0.70) {MPSGE.jl};
  \node[chip, text width=1.7cm] at (-1.18,0.70) {pycge};
  \node[chip, text width=2.1cm] at ( 1.05,0.70) {MIRAGRODEP};
  \node[chip, text width=1.7cm] at ( 3.27,0.70) {OpenCGE};
  \node[chip, text width=2.8cm] at ( 5.84,0.70) {OSeMOSYS\\MESSAGEix\,$\cdot$\,TEMOA};
  \draw[flow] (combat.south) -- (com.north);
  \node[fwd] at ($(combat.south)!0.5!(com.north)$) {physical disruption parameters};
  \draw[flow] (com.south) -- (down.north);
  \node[fwd] at ($(com.south)!0.5!(down.north)$) {helium sectoral shortfall};
  \draw[flow] (down.south) -- (srm.north);
  \node[fwd] at ($(down.south)!0.5!(srm.north)$) {commodity price \& supply shock vector};
  \draw[flow] (srm.south) -- (lrm.north);
  \node[fwd] at ($(srm.south)!0.5!(lrm.north)$) {structural-layer calibration};
  \def\lane{-7.5}
  \coordinate (bpo) at ([yshift=-0.60cm]com.west);
  \coordinate (bpi) at ([yshift=0.55cm]lrm.west);
  \draw[-{Stealth[length=2.4mm]}, dashed, line width=0.9pt, amber, rounded corners=4pt]
        (bpo) -- (bpo-|\lane,0) -- (bpi-|\lane,0) -- (bpi);
  \node[font=\sffamily\scriptsize\itshape, amber, rotate=90, anchor=south]
        at ($(bpo-|\lane,0)!0.5!(bpi-|\lane,0)+(-0.22,0)$)
        {water unmet-demand (Scenario~E only)};
  \node[draw=slate, line width=0.8pt, rounded corners=3pt, fill=slate!8,
        minimum width=1.9cm, minimum height=10.75cm, align=center,
        font=\sffamily\bfseries\small, anchor=center] (strat) at (9.05,5.325)
        {\rotatebox{90}{Strategic assessment \;\textnormal{\itshape(parallel synthesis)}}};
  \foreach \b in {combat,com,down,srm,lrm}{
    \draw[-{Stealth[length=2mm]}, line width=0.7pt, slate, dotted]
          (\b.east) -- (\b.east-|strat.west);
  }
  \node[font=\sffamily\footnotesize\itshape, slate, anchor=north west, text width=8.5cm]
        at (-4.4,-0.45)
        {Models within a level run in parallel; levels run in sequence so no model executes before its upstream inputs exist. Models that cannot run are recorded as skipped, not fatal.};
\end{tikzpicture}}
\caption{The analytical-level stack and the flow of information across it. Domain models are grouped into a combat layer, a commodity layer for the five most-affected markets, a commodity-downstream layer whose models consume other commodity outputs, a short-run macroeconomic layer, and a long-run macroeconomic and energy-systems layer; a strategic-assessment track synthesizes from all levels in parallel. Solid arrows carry quantities forwarded automatically from upstream model output to downstream model input (Methods). The dashed amber path is a scenario-scoped rule that routes a water unmet-demand signal into the macroeconomic layer only under the infrastructure-collapse scenario (Scenario~E). Models shown are illustrative of each layer; Appendix~\ref{secA1} gives the full inventory and execution status.}
\label{fig:level_stack}
\end{figure*}
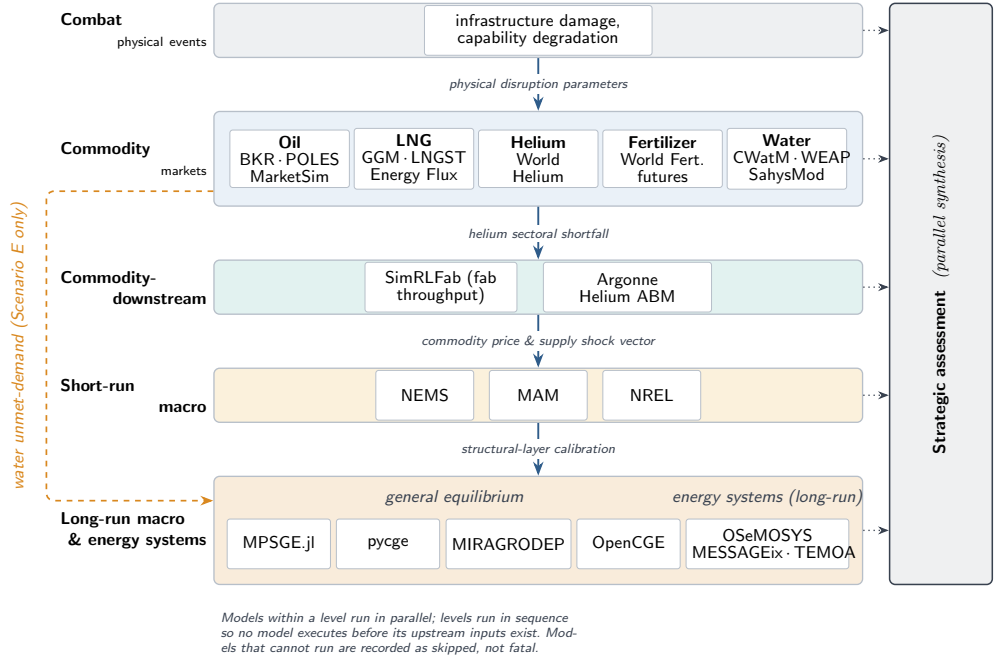

The portfolio is organized into analytical levels: a combat layer, a commodity layer for the five markets, a commodity-downstream layer for models that consume other commodity outputs (for example, the semiconductor-fabrication model consumes the helium model's sectoral shortfall), a short-run macroeconomic layer, and a long-run macroeconomic and energy-systems layer, with a strategic-assessment track running in parallel (Fig.~\ref{fig:level_stack}). Models within a level run concurrently; levels run in sequence, so no model executes before its upstream inputs exist.

\section{Reading the crisis across scenario, region, and time}\label{sec:results}

The framework returns an artifact rather than a point estimate: a synthesis that states each magnitude, interprets it, and flags where the models disagree. Comparison views then read the same crisis across scenarios, by region, and over time. Every number is attributed to a named model under a named scenario. Because most models in this run operate as literature-calibrated reductions, the figures are forward-looking projections rather than forecasts.

The demonstration run orchestrated sixteen of the thirty-four registered models across the five scenarios. Methods describes the full registry of adapters. The remainder awaited upstream code, licensed binaries, or the required run configuration and were recorded as skipped; the pipeline is designed to continue when individual models cannot run.

Coverage varies by scenario, and coverage determines how much weight a comparison can bear. Scenario C drew on all sixteen models, Scenario B on eleven, Scenario A on eight, and Scenarios D and E on five each. Five anchor models ran under every scenario. A comparison resting on sixteen models does not carry the same evidential weight as one resting on five, and the coverage record lets the analyst weigh accordingly. Appendix~\ref{secA4} gives the full coverage map.

\subsection{Across scenarios}\label{subsec:across}

\begin{figure*}[t]
\centering
\begin{adjustbox}{max width=\textwidth}
\begin{tikzpicture}
\begin{groupplot}[
    group style={group size=3 by 2, horizontal sep=1.6cm, vertical sep=2.0cm},
    width=0.30\textwidth, height=5.2cm,
    symbolic x coords={A,B,C,D,E}, xtick=data, enlarge x limits=0.13,
    tick align=outside, tickpos=left,
    xlabel={Scenario}, xlabel style={font=\footnotesize}, ylabel style={font=\footnotesize},
    ylabel near ticks,
    xticklabel style={font=\footnotesize}, yticklabel style={font=\footnotesize},
    ymajorgrids, grid style={linegray!55, line width=0.3pt},
    extra y ticks={0}, extra y tick style={grid=major, grid style={slate, line width=0.5pt}},
    title style={font=\footnotesize\bfseries, anchor=south, yshift=1pt, align=center, text width=3.4cm},
]
\nextgroupplot[title={(a) Oil impulse}, ymin=-2, ymax=70, ylabel={\%}]
    \addplot[inkblue, line width=0.7pt, mark=*, mark size=1.7pt, mark options={fill=inkblue}] coordinates {(A,54.79) (B,60.5) (C,58.25) (D,51.82) (E,48.36)};
    \addplot[scAm, line width=0.7pt, mark=diamond*, mark size=1.7pt, mark options={fill=scAm}] coordinates {(A,54.79) (C,58.25)};
    \addplot[scStl, line width=0.7pt, mark=square*, mark size=1.7pt, mark options={fill=scStl}] coordinates {(A,54.79) (B,60.5) (C,58.25) (D,51.82) (E,48.36)};
\nextgroupplot[title={(b) LNG / gas\,$\ddagger$}, ymin=-5, ymax=140, ylabel={\%}]
    \addplot[scLt, line width=0.7pt, mark=*, mark size=1.7pt, mark options={fill=scLt}] coordinates {(B,50.926) (C,50.926)};
    \addplot[slate, line width=0.7pt, mark=x, mark size=1.7pt, mark options={fill=slate}] coordinates {(B,21.0) (C,21.0)};
    \addplot[scRd, line width=0.7pt, mark=pentagon*, mark size=1.7pt, mark options={fill=scRd}] coordinates {(A,50.92727272727273) (B,126.39090909090908) (C,126.39090909090908) (D,126.39090909090908) (E,126.39090909090908)};
\nextgroupplot[title={(c) GDP}, ymin=-1.7, ymax=0.15, ylabel={\%}]
    \addplot[inkblue, line width=0.7pt, mark=*, mark size=1.7pt, mark options={fill=inkblue}] coordinates {(A,-0.6164) (B,-0.6806) (C,-0.6553) (D,-0.8258) (E,-1.5415)};
    \addplot[scStl, line width=0.7pt, mark=square*, mark size=1.7pt, mark options={fill=scStl}] coordinates {(A,-0.025) (B,-0.8883) (C,-0.8695) (D,-1.4314) (E,-1.0355)};
    \addplot[scTl, line width=0.7pt, mark=triangle*, mark size=1.7pt, mark options={fill=scTl}] coordinates {(A,-0.157) (B,-0.8139) (C,-0.7889) (D,-1.0575) (E,-0.8636)};
\nextgroupplot[title={(d) CPI / inflation}, ymin=0, ymax=3.3, ylabel={pp \,/\, \%}]
    \addplot[inkblue, line width=0.7pt, mark=*, mark size=1.7pt, mark options={fill=inkblue}] coordinates {(A,1.0958) (B,1.21) (C,1.165) (D,1.101) (E,2.0553)};
    \addplot[scStl, line width=0.7pt, mark=square*, mark size=1.7pt, mark options={fill=scStl}] coordinates {(A,1.6331) (B,2.7989) (C,2.7539) (D,3.0628) (E,2.6609)};
    \addplot[scTl, line width=0.7pt, mark=triangle*, mark size=1.7pt, mark options={fill=scTl}] coordinates {(A,1.3458) (B,1.757) (C,1.712) (D,2.0668) (E,1.7945)};
\nextgroupplot[title={(e) Fleet loss}, ymin=-3, ymax=35, ylabel={\%}]
    \addplot[scRust, line width=0.7pt, mark=square*, mark size=1.7pt, mark options={fill=scRust}] coordinates {(B,31.154) (C,31.154)};
    \addplot[slate, line width=0.7pt, mark=o, mark size=1.7pt, mark options={fill=slate}] coordinates {(B,0.0) (C,0.0)};
\end{groupplot}
\end{tikzpicture}
\end{adjustbox}
\par\vspace{1.2mm}
\begin{tikzpicture}
\begin{axis}[hide axis, scale only axis, width=1mm, height=1mm,
    xmin=0,xmax=1,ymin=0,ymax=1,
    legend style={draw=linegray, line width=0.3pt, font=\scriptsize, fill=white,
        at={(0,0)}, anchor=center, legend columns=5,
        /tikz/every even column/.append style={column sep=5pt},
        column sep=5pt},
    legend cell align=left]
\addlegendimage{inkblue, line width=0.7pt, mark=*, mark size=1.7pt, mark options={fill=inkblue}}\addlegendentry{Fed oil}
\addlegendimage{scStl, line width=0.7pt, mark=square*, mark size=1.7pt, mark options={fill=scStl}}\addlegendentry{MPSGE.jl}
\addlegendimage{scTl, line width=0.7pt, mark=triangle*, mark size=1.7pt, mark options={fill=scTl}}\addlegendentry{pycge}
\addlegendimage{scAm, line width=0.7pt, mark=diamond*, mark size=1.7pt, mark options={fill=scAm}}\addlegendentry{POLES-JRC}
\addlegendimage{scLt, line width=0.7pt, mark=*, mark size=1.7pt, mark options={fill=scLt}}\addlegendentry{LNGST (LNG)}
\addlegendimage{slate, line width=0.7pt, mark=x, mark size=1.7pt, mark options={fill=slate}}\addlegendentry{Fwd-curve}
\addlegendimage{scRd, line width=0.7pt, mark=pentagon*, mark size=1.7pt, mark options={fill=scRd}}\addlegendentry{Nat-gas (NREL / CGE)}
\addlegendimage{scRust, line width=0.7pt, mark=square*, mark size=1.7pt, mark options={fill=scRust}}\addlegendentry{AISdb}
\addlegendimage{slate, line width=0.7pt, mark=o, mark size=1.7pt, mark options={fill=slate}}\addlegendentry{AIS-alt}
\end{axis}
\end{tikzpicture}
\caption{Reading impacts across scenarios. Small multiples of five headline quantities across Scenarios A to E, one marker per contributing model: (a) oil-price impulse; (b) the LNG and natural-gas signal; (c) GDP impact; (d) CPI and inflation; (e) effective shipping-fleet-capacity loss. The LNG and natural-gas signal ($\ddagger$) is the quantity the cross-model consistency pass flags, because the LNG, forward-curve, and electricity/CGE series price three different objects and are not reconcilable as a single number. Panel (e) displays the unreconciled shipping disagreement as a visible spread rather than an average. All points are passed-through model outputs.}
\label{fig:results_scenario}
\end{figure*}
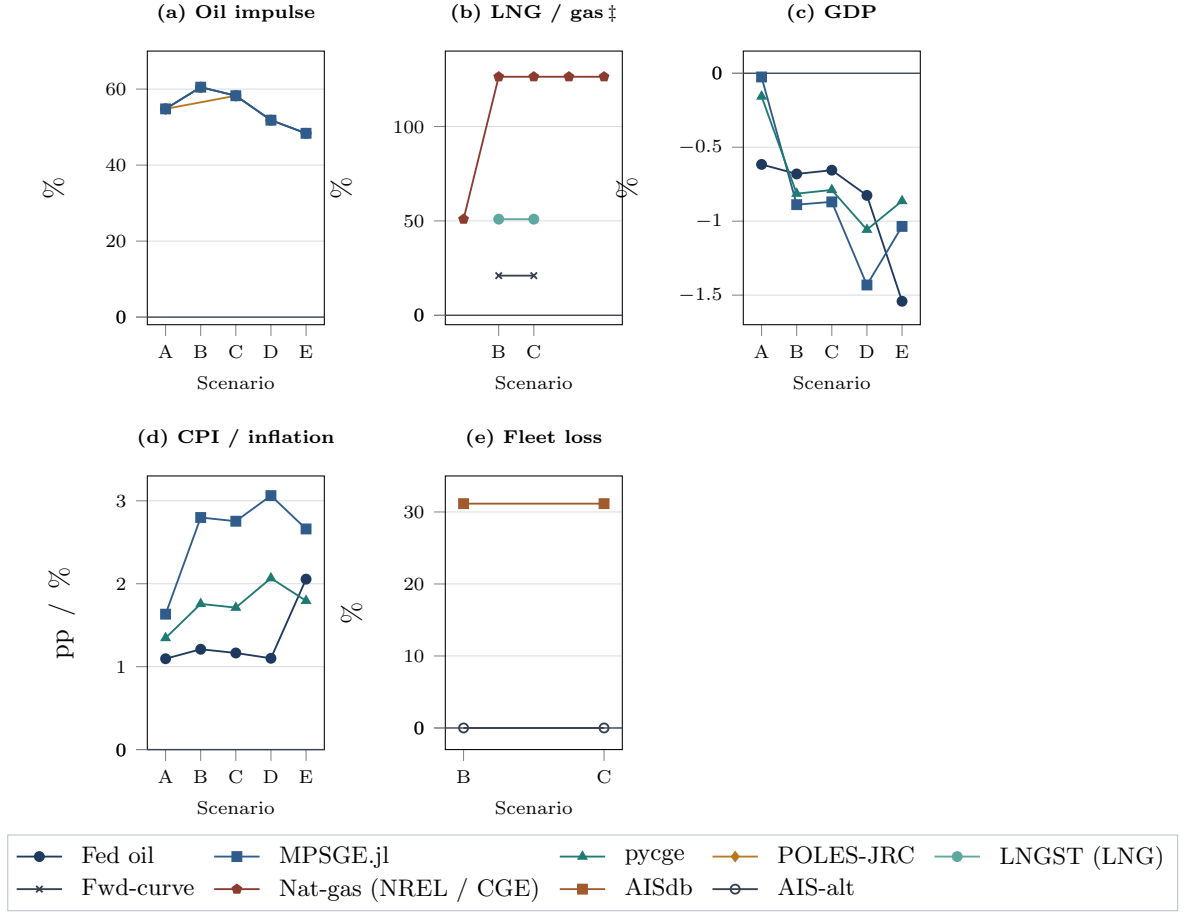

Placed side by side, the models that generate a given quantity reveal where the portfolio agrees and where it diverges beyond declared tolerances (Fig.~\ref{fig:results_scenario}). Three features stand out.

First, the oil-price impulse is not monotone in severity. Across Scenarios A to E the Federal Reserve oil model \cite{bib_fed_oil} reported impulses of $+54.8\%$, $+60.5\%$, $+58.3\%$, $+51.8\%$, and $+48.4\%$. The largest impulse arises under the prolonged-but-contained Scenario B, not the tail. An analyst working one scenario at a time could miss that the worst price impulse does not coincide with the worst scenario.

Second, agreement can be mechanical. Under Scenario C the $+58.3\%$ oil shock appears identically in the Federal Reserve model, in POLES-JRC \cite{bib_poles}, and in both CGE models, because the macro layer received the commodity-tier price rather than re-deriving one from the narrative. Agreement of this kind reflects the pipeline's linkage rules, not independent confirmation.

Third, the gas signal is where the portfolio genuinely diverges, and the synthesis states the disagreement rather than reporting an average. Under Scenarios B and C the LNG spreadsheet tool reported a price impulse near $+50.9\%$, a forward-curve model reported $+21\%$, and the electricity and CGE models carried a natural-gas price increase of $+126.4\%$. These series price three different objects: the delivered-LNG price, the futures forward curve, and the domestic hub price. They cannot be reconciled into one number, so the consistency pass flags the spread and directs the analyst to the substantive question underneath, prompting an assessment of whether the divergence reflects the complexity of the system or differences in model assumptions. A starker case: two shipping models of the same quantity reported effective fleet-capacity losses of $31.2\%$ and $0\%$, a disagreement preserved rather than averaged.

One Scenario-A outlier illustrates the discipline directly. Under the swift-and-contained scenario, the pycge model reported a global GDP impact of $-15.7\%$, one to two orders of magnitude larger than every other model of the same scenario and out of line with pycge's own outputs under the more severe scenarios. The number is kept, flagged, and traceable to the run that produced it, so the analyst sees the value and the warning together. Figures~\ref{fig:results_scenario} and~\ref{fig:results_regional} plot the value from a subsequent analyst-approved selective re-run ($-0.157\%$); the flagged outlier remains in the provenance record of the original run.

\subsection{By region}\label{subsec:region}

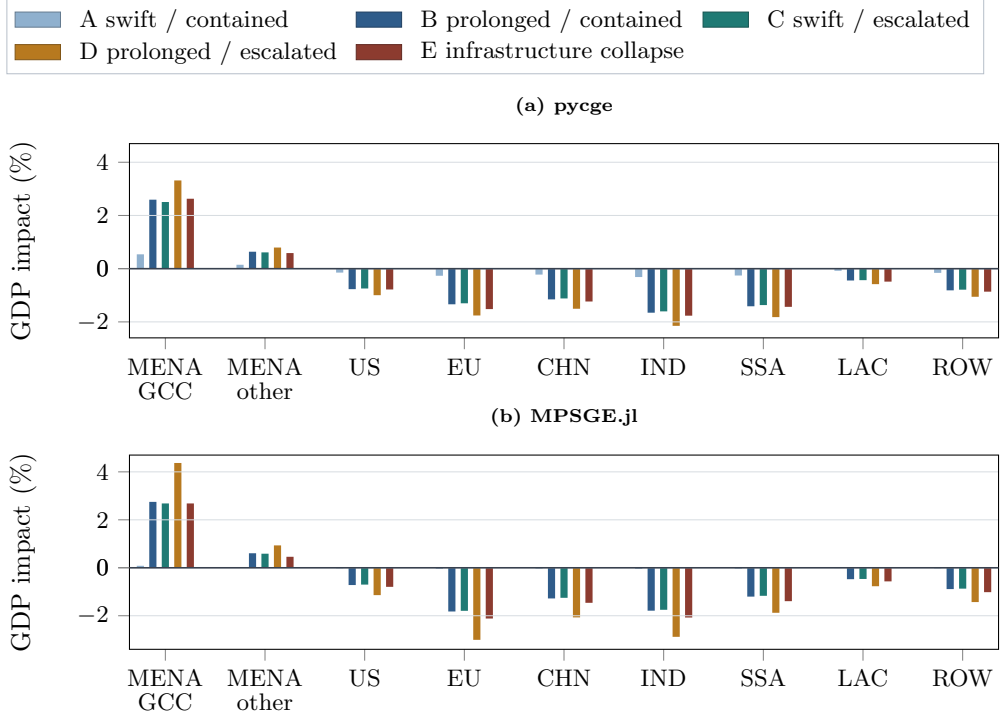
\begin{figure*}[t]
\centering
\begin{tikzpicture}
\begin{axis}[hide axis, scale only axis, width=1mm, height=1mm,
    xmin=0,xmax=1,ymin=0,ymax=1,
    legend style={draw=linegray, line width=0.3pt, font=\scriptsize, fill=white,
        at={(0,0)}, anchor=center, legend columns=3,
        /tikz/every even column/.append style={column sep=5pt},
        column sep=5pt},
    legend cell align=left]
\addlegendimage{area legend, fill=scA, draw=scA!70!black, line width=0.2pt}\addlegendentry{A  swift / contained}
\addlegendimage{area legend, fill=scB, draw=scB!70!black, line width=0.2pt}\addlegendentry{B  prolonged / contained}
\addlegendimage{area legend, fill=scC, draw=scC!70!black, line width=0.2pt}\addlegendentry{C  swift / escalated}
\addlegendimage{area legend, fill=scD, draw=scD!70!black, line width=0.2pt}\addlegendentry{D  prolonged / escalated}
\addlegendimage{area legend, fill=scE, draw=scE!70!black, line width=0.2pt}\addlegendentry{E  infrastructure collapse}
\end{axis}
\end{tikzpicture}
\par\vspace{1.5mm}
\begin{tikzpicture}
\begin{groupplot}[
    group style={group size=1 by 2, vertical sep=1.55cm},
    width=\textwidth, height=4.15cm,
    enlarge x limits=0.045,
    xtick={0,1,2,3,4,5,6,7,8}, xticklabels={{MENA\\GCC},{MENA\\other},{US},{EU},{CHN},{IND},{SSA},{LAC},{ROW}},
    xticklabel style={font=\scriptsize, align=center, text width=1.35cm},
    ymajorgrids, grid style={linegray!55, line width=0.3pt},
    axis on top, tick align=outside, tickpos=left,
    every axis plot/.append style={line width=0.2pt},
    ylabel style={font=\scriptsize}, yticklabel style={font=\scriptsize},
    ylabel near ticks,
    extra y ticks={0}, extra y tick style={grid=major, grid style={slate, line width=0.6pt}},
    title style={font=\footnotesize\bfseries, anchor=south, yshift=1pt},
]
\nextgroupplot[ybar, bar width=2.7pt, ylabel={GDP impact (\%)}, title={(a) pycge}, ymin=-2.6, ymax=4.7]
    \addplot+[ybar, draw=none, fill=scA] coordinates {(0,0.5394) (1,0.147) (2,-0.147) (3,-0.2655) (4,-0.2218) (5,-0.314) (6,-0.2566) (7,-0.0785) (8,-0.157)};
    \addplot+[ybar, draw=none, fill=scB] coordinates {(0,2.5929) (1,0.6374) (2,-0.7694) (3,-1.3383) (4,-1.1536) (5,-1.6541) (6,-1.4127) (7,-0.4439) (8,-0.8139)};
    \addplot+[ybar, draw=none, fill=scC] coordinates {(0,2.5054) (1,0.6124) (2,-0.7444) (3,-1.3008) (4,-1.1186) (5,-1.6041) (6,-1.3677) (7,-0.4314) (8,-0.7889)};
    \addplot+[ybar, draw=none, fill=scD] coordinates {(0,3.3161) (1,0.7945) (2,-0.9962) (3,-1.7588) (4,-1.5056) (5,-2.1479) (6,-1.8198) (7,-0.5811) (8,-1.0555)};
    \addplot+[ybar, draw=none, fill=scE] coordinates {(0,2.6286) (1,0.5866) (2,-0.7814) (3,-1.5186) (4,-1.2333) (5,-1.7661) (6,-1.4341) (7,-0.4863) (8,-0.8636)};
\nextgroupplot[ybar, bar width=2.7pt, ylabel={GDP impact (\%)}, title={(b) MPSGE.jl}, ymin=-3.4, ymax=4.7]
    \addplot+[ybar, draw=none, fill=scA] coordinates {(0,0.0832) (1,0.0215) (2,-0.0222) (3,-0.0458) (4,-0.0354) (5,-0.0497) (6,-0.0374) (7,-0.0127) (8,-0.025)};
    \addplot+[ybar, draw=none, fill=scB] coordinates {(0,2.7484) (1,0.6058) (2,-0.7198) (3,-1.8224) (4,-1.2783) (5,-1.7889) (6,-1.2023) (7,-0.4756) (8,-0.8883)};
    \addplot+[ybar, draw=none, fill=scC] coordinates {(0,2.6828) (1,0.5871) (2,-0.701) (3,-1.7942) (4,-1.252) (5,-1.7514) (6,-1.1685) (7,-0.4662) (8,-0.8695)};
    \addplot+[ybar, draw=none, fill=scD] coordinates {(0,4.3702) (1,0.9336) (2,-1.14) (3,-3.0045) (4,-2.0685) (5,-2.8791) (6,-1.8761) (7,-0.769) (8,-1.4297)};
    \addplot+[ybar, draw=none, fill=scE] coordinates {(0,2.6831) (1,0.4588) (2,-0.7945) (3,-2.1161) (4,-1.4603) (5,-2.0742) (6,-1.392) (7,-0.5666) (8,-1.0165)};
\end{groupplot}
\end{tikzpicture}
\caption{Reading impacts by region. GDP impact (\%) by region under each scenario, for the two general-equilibrium models that resolve regional detail (pycge, top; MPSGE.jl, bottom). Regions run from the Gulf-producer aggregate (MENA\_GCC) through the major importers to the rest of the world (ROW). Both models reproduce the same producer-versus-importer asymmetry, the Gulf aggregate gaining while import-dependent regions lose, and differ chiefly in magnitude. The remaining models returned a single global figure with no regional dispersion and so do not appear here. All bars are passed-through model outputs.}
\label{fig:results_regional}
\end{figure*}

Regional resolution comes mainly from the two CGE models, supplemented by the water and fertilizer models. Most other models return a single global figure, which the synthesis records rather than hides (Fig.~\ref{fig:results_regional}).

Where the CGE models resolve regions, they agree on direction and differ on magnitude. Under Scenario C, MPSGE.jl reported a global GDP impact near $-0.87\%$ alongside a gain of about $+2.68\%$ for the Gulf-producer aggregate. pycge reported $-0.79\%$ globally, a comparable Gulf gain of $+2.51\%$, and the sharpest loss in Sub-Saharan Africa at $-1.37\%$.

The producer-versus-importer asymmetry recurs. Under Scenario D pycge reported a Gulf gain above $+3\%$ against an Indian-aggregate loss beyond $-2\%$. The fertilizer model adds a matching trade reallocation: a 30\% loss of Middle Eastern production is redistributed as higher exports from the United States ($+6\%$), Russia ($+8.4\%$), and Algeria ($+4.8\%$).

Water enters the regional view only under escalation. Under Scenario C the regional water models \cite{bib_weap,bib_watergap} registered what the contained scenarios leave dormant: a 30\% loss of desalination capacity, unmet demand of 9--13\%, and roughly 30 million people affected, with the UAE and Qatar most exposed. One coupling rule did not fire in this run.

\begin{figure}[t]
    \centering
    \includegraphics[width=0.9\linewidth]{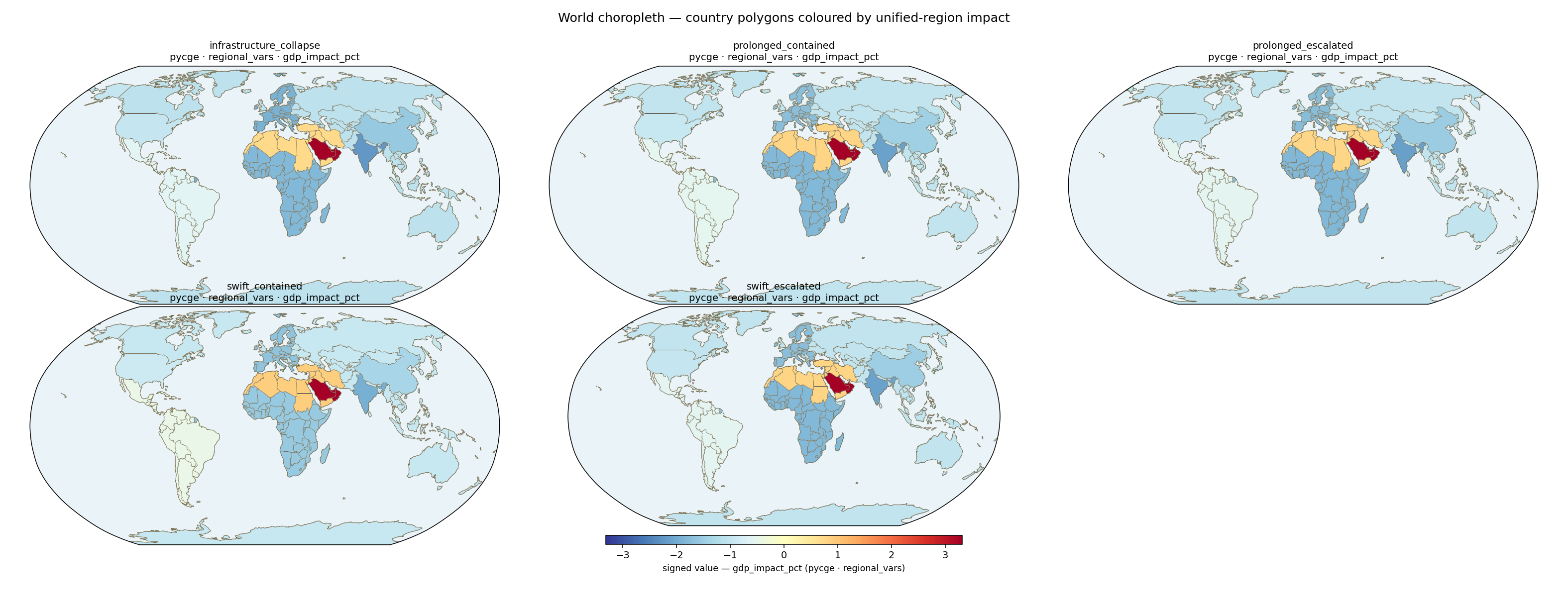}
    \caption{Regional disparities in economic impact. A choropleth of GDP impact by region under the scenarios. Higher oil prices register as a net GDP gain for the MENA producer aggregate in most scenarios, despite the accompanying challenges to regional stability.}
    \label{fig:Mapping}
\end{figure}

The rule that forwards water unmet demand into the macroeconomic layer is scoped to Scenario~E; no water model was dispatched under that scenario, so the rule found no upstream value and the framework fell back to the LLM-extracted water shock, its standard behavior when an upstream source is absent (Methods). Geographic disparities in the GDP response are summarized in Fig.~\ref{fig:Mapping}.

\subsection{Over time}\label{subsec:time}

\begin{figure*}[t]
\centering
\begin{tikzpicture}
\begin{groupplot}[
    group style={group size=2 by 1, horizontal sep=1.7cm},
    width=0.52\textwidth, height=5.4cm,
    tick align=outside, tickpos=left,
    title style={font=\footnotesize\bfseries, anchor=south, yshift=1pt},
    ylabel style={font=\scriptsize}, xlabel style={font=\scriptsize},
    yticklabel style={font=\scriptsize}, xticklabel style={font=\scriptsize},
]
\nextgroupplot[
    title={(a) Federal Reserve model: peak impacts},
    ybar, bar width=4.2pt, enlarge x limits=0.14,
    symbolic x coords={A,B,C,D,E}, xtick=data,
    xticklabels={{A\\{\tiny 1\,Q}},{B\\{\tiny 1\,Q}},{C\\{\tiny 1\,Q}},{D\\{\tiny 1.75\,Q}},{E\\{\tiny 5\,Q}}}, xticklabel style={font=\scriptsize, align=center},
    xlabel={Scenario \;(elevated-price duration)}, ylabel={Impact},
    ymin=-2.0, ymax=2.45, ymajorgrids, grid style={linegray!55, line width=0.3pt},
    extra y ticks={0}, extra y tick style={grid=major, grid style={slate, line width=0.6pt}},
    legend style={font=\tiny, at={(0.5,1.23)}, anchor=south, legend columns=4,
                  draw=linegray, /tikz/every even column/.append style={column sep=4pt}},
]
    \addplot+[ybar, draw=none, fill=scC] coordinates {(A,-0.6164) (B,-0.6806) (C,-0.6553) (D,-0.8258) (E,-1.5415)};
    \addplot+[ybar, draw=none, fill=scD] coordinates {(A,1.0958) (B,1.21) (C,1.165) (D,1.101) (E,2.0553)};
    \addplot+[ybar, draw=none, fill=scB] coordinates {(A,0.3082) (B,0.3403) (C,0.3277) (D,0.4129) (E,0.7708)};
    \addplot+[ybar, draw=none, fill=slate] coordinates {(A,0.5479) (B,0.605) (C,0.5825) (D,0.5505) (E,1.0276)};
\legend{Peak GDP (\%), Peak CPI (pp), Peak unemp. (pp), Peak FFR (pp)}
\nextgroupplot[
    title={(b) Energy~Flux: cumulative gas-power build},
    axis y line*=left, xlabel={Year}, ylabel={Added capacity (GW)},
    xmin=2025.5, xmax=2035.5, ymin=0, ymax=110,
    xtick={2026,2028,2030,2032,2034}, ymajorgrids, grid style={linegray!55, line width=0.3pt},
    scaled x ticks=false, x tick label style={/pgf/number format/1000 sep={}},
]
\addplot[scC, line width=1.1pt, mark=*, mark size=1.3pt] coordinates {(2026,10.135) (2027,20.27) (2028,30.405) (2029,40.54) (2030,50.675) (2031,60.81) (2032,70.945) (2033,81.08) (2034,91.215) (2035,101.35)};
\node[anchor=north west, font=\tiny, align=left, inner sep=1.5pt] at (axis description cs:0.04,0.97)
  {\textcolor{scC}{\rule[0.45ex]{9pt}{1.1pt}}~capacity (left)\\[-0.5pt]
   \textcolor{amber}{$\blacklozenge$}~deliverability (right)};
\end{groupplot}
\begin{axis}[
    width=0.52\textwidth, height=5.4cm,
    at={(group c2r1.south east)}, anchor=south east,
    axis y line*=right, axis x line=none,
    xmin=2025.5, xmax=2035.5, ymin=0, ymax=8.3,
    ylabel={Deliverability (Bcf/d)}, ylabel style={font=\scriptsize},
    yticklabel style={font=\scriptsize}, ytick={0,2,4,6,8},
]
\addplot[amber, only marks, mark=diamond*, mark size=2.4pt] coordinates {(2030,3.799891996142719) (2035,7.599783992285438)};
\end{axis}
\end{tikzpicture}
\caption{Reading impacts over time. (a) Peak macroeconomic impacts from the Federal Reserve oil model under each scenario, with the duration of elevated prices (in quarters) printed beneath each scenario label. (b) The Energy~Flux cumulative gas-power build-out, 2026 to 2035: added capacity on the left axis and the model's two reported incremental-deliverability points on the right axis. The panels are shown on their native resolutions rather than interpolated to a common grid. All values are passed-through model outputs under the indicated scenarios.}
\label{fig:results_temporal}
\end{figure*}
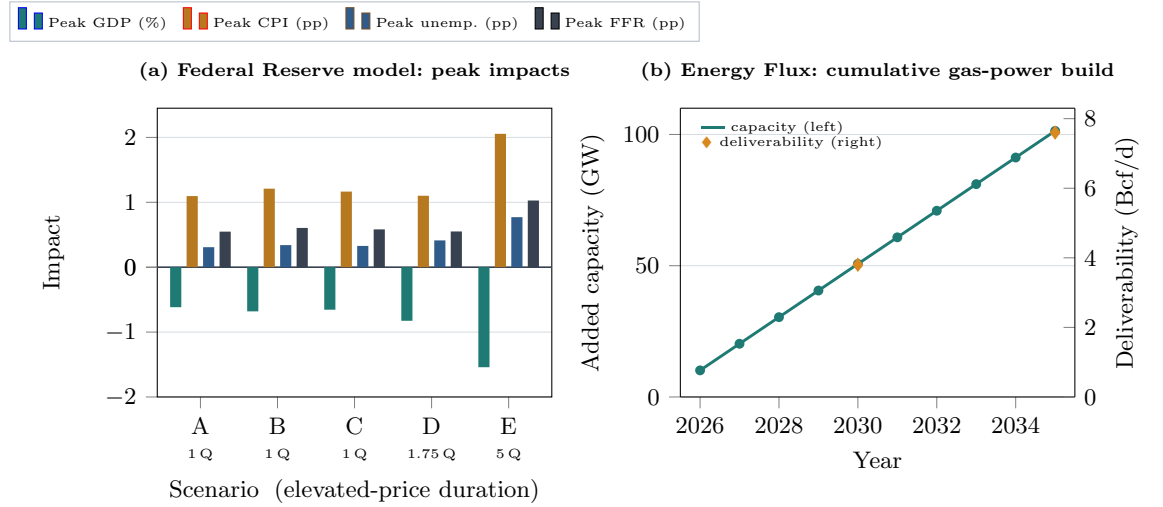

Three models carry an explicit time index and resolve different horizons, so the temporal view is genuinely multi-resolution (Fig.~\ref{fig:results_temporal}). The Federal Reserve model returns quarterly macroeconomic paths: the peak GDP impact deepens from about $-0.62\%$ under Scenario A to $-1.54\%$ under the tail, and the duration of elevated prices lengthens from one quarter to roughly five. The Energy Flux build-out model returns an annual trajectory, with cumulative added gas-power capacity rising from about 10~GW in 2026 to about 101~GW by 2035.

The same view records what is robust. The helium market is reported as essentially undisrupted across all scenarios, its equilibrium price held at the \$280 per thousand standard cubic feet baseline with aerospace and defense demand rationed first \cite{bib_argonne_helium}, and the fertilizer index sits near $+23.8\%$ wherever it is resolved. Robustness and fragility appear together, which is the point.

\section{Refreshing the analysis}\label{sec:temporal}

The framework can be refreshed along two axes: as the information set changes from week to week, and as the orchestrating model itself is replaced. We exercise both.

\subsection{As new information arrives}\label{subsec:temporal_info}

A protracted chokepoint crisis is not a single discrete shock. Parameters defensibly treated as uncertain at onset become informative priors weeks later, so a framework that claims to lower the cost of multi-model analysis should be judged on whether it can refresh a reading as the information set changes.

We therefore add a temporal layer that re-runs the entire pipeline weekly, producing a sequence of reports indexed by the week of information they rest on. For the demonstration we instantiated it as eight weekly re-runs over the window from 15 February to 10 April 2026 (week-start dates 15 February through 5 April), bracketing the onset and initial stabilization of the closure.
Each week draws on commercial news feeds and US Energy Information Administration indicator series, which an LLM summarizer maps into a structured weekly brief that may arrange source-attested quantities but may not originate numbers.

\begin{figure*}[t]
\centering
\begin{tikzpicture}
\begin{groupplot}[
    group style={group size=2 by 1, horizontal sep=1.9cm},
    width=0.52\textwidth, height=5.4cm,
    symbolic x coords={Feb15,Feb22,Mar01,Mar08,Mar15,Mar22},
    xtick=data, enlarge x limits=0.10,
    tick align=outside, tickpos=left,
    title style={font=\footnotesize\bfseries, anchor=south, yshift=1pt},
    ylabel style={font=\scriptsize}, xlabel style={font=\scriptsize},
    yticklabel style={font=\scriptsize},
    xticklabel style={font=\scriptsize, rotate=32, anchor=east},
    ymajorgrids, grid style={linegray!55, line width=0.3pt},
]
\nextgroupplot[
    title={(a) Commodity and shipping stress},
    ylabel={\%}, ymin=0, ymax=98,
    legend style={font=\tiny, at={(0.5,1.24)}, anchor=south, legend columns=2,
                  draw=linegray, /tikz/every even column/.append style={column sep=5pt}},
]
\addplot[inkblue, line width=0.9pt, mark=*, mark size=1.6pt] coordinates {(Feb15,56.485) (Feb22,75.312) (Mar01,59.882) (Mar08,63.336) (Mar15,59.388) (Mar22,83.347)};
\addplot[scRust, line width=0.9pt, mark=square*, mark size=1.6pt] coordinates {(Feb15,40.177) (Feb22,28.806) (Mar01,30.421) (Mar08,30.421) (Mar15,25.177) (Mar22,22.384)};
\legend{oil-price impulse, fleet-capacity loss}
\node[anchor=south, font=\tiny\itshape, slate, align=center] at (axis cs:Feb22,79) {Epic Fury\\(28 Feb)};
\node[anchor=south, font=\tiny\itshape, slate, align=center] at (axis cs:Mar22,86.5) {Trump\\ultimatum};
\nextgroupplot[
    title={(b) Macro response and information set},
    ylabel={GDP (\%) \,/\, CPI (pp)}, ymin=-2.2, ymax=3.2,
    extra y ticks={0}, extra y tick style={grid=major, grid style={slate, line width=0.6pt}},
    legend style={font=\tiny, at={(0.5,1.24)}, anchor=south, legend columns=2,
                  draw=linegray, /tikz/every even column/.append style={column sep=5pt}},
]
\addplot[scStl, line width=0.9pt, mark=triangle*, mark size=1.9pt] coordinates {(Feb15,-1.064) (Feb22,-1.325) (Mar01,-0.893) (Mar08,-1.296) (Mar15,-0.573) (Mar22,-1.245)};
\addplot[scTl, line width=0.9pt, mark=diamond*, mark size=1.9pt] coordinates {(Feb15,1.894) (Feb22,2.511) (Mar01,1.762) (Mar08,2.394) (Mar15,1.480) (Mar22,2.582)};
\legend{GDP impact, CPI inflation}
\end{groupplot}
\begin{axis}[
    width=0.52\textwidth, height=5.4cm,
    at={(group c2r1.south east)}, anchor=south east,
    axis y line*=right, axis x line=none,
    symbolic x coords={Feb15,Feb22,Mar01,Mar08,Mar15,Mar22},
    enlarge x limits=0.10,
    ymin=0, ymax=340, ylabel={articles ingested (coverage)},
    ylabel style={font=\scriptsize}, yticklabel style={font=\scriptsize},
]
\addplot[amber, line width=0.8pt, densely dashed, mark=o, mark size=1.5pt] coordinates {(Feb15,196) (Feb22,226) (Mar01,243) (Mar08,273) (Mar15,275) (Mar22,266)};
\node[anchor=north east, font=\tiny, align=right, inner sep=1.5pt] at (axis description cs:0.97,0.97)
  {\textcolor{amber}{\rule[0.4ex]{7pt}{0.8pt}}~articles (right)};
\end{axis}
\end{tikzpicture}
\caption{Refreshing the analysis over time. Headline quantities from the first six of the eight consecutive weekly re-runs (15 February to 22 March 2026), each executed under the \texttt{gpt-oss-120b} orchestration backend on that week's news-derived crisis description with the scenario definitions, model registry, forwarding rules, and consistency tolerances held fixed. (a) The oil-price impulse and effective shipping-fleet-capacity loss, both aggregated passed-through model outputs. (b) The GDP and CPI responses (left axis) against the number of source articles ingested that week (right axis, dashed), so that a movement driven by a genuine information change is distinguishable from one driven by degraded coverage. The oil-price impulse is non-monotone in calendar time, peaking in the week of the 22 March Hormuz ultimatum rather than at closure onset. Because everything except the information set is held fixed, any week-to-week change resolves through the provenance chain either to a change in extracted parameters or to a change in the pipeline's behavior under identical inputs. The view turns a single cross-sectional reading into a monitored trajectory.}
\label{fig:results_weekly}
\end{figure*}
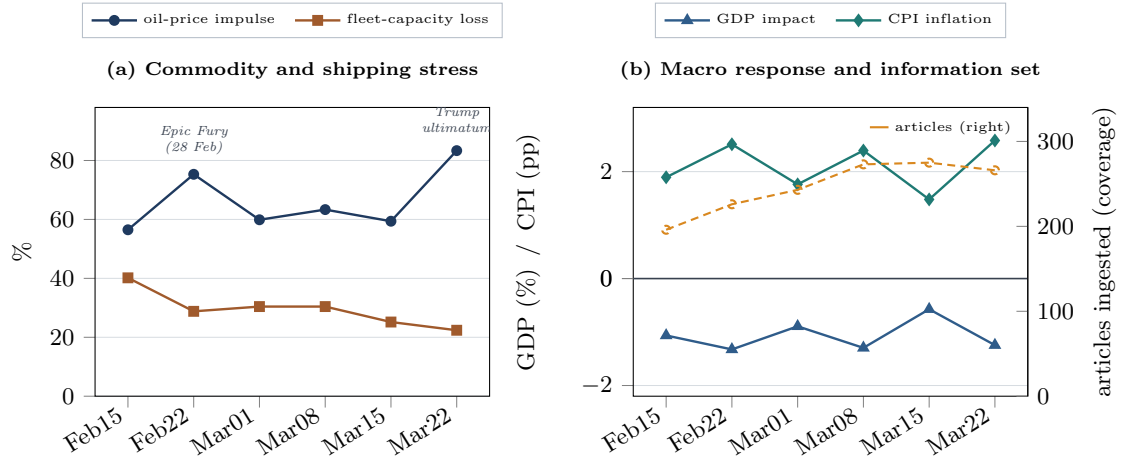

Because scenario definitions, the model registry, the forwarding rules, and the consistency tolerances are held fixed, successive runs are directly comparable. Any week-to-week change resolves through the provenance chain either to a change in extracted parameters or to a change in the pipeline's behavior under identical inputs, the latter flagged as a regression. In effect, each refresh performs a form of data assimilation and recalibration: as new data and knowledge arrive, the framework updates the distribution of feasible outcomes---conditional on the set of models selected and on how the situation has evolved---while retaining the historical memory of prior weeks. The results the analyst refreshes are thus a controlled re-measurement rather than a fresh guess (Fig.~\ref{fig:results_weekly}).

What the analyst reads first each week is the summarizer's brief, and the sequence of briefs is itself a decision-relevant object. The headlines of three briefs from the sequence convey the trajectory a policymaker would follow. Briefs are labelled by the first day of the seven-day window they cover, so the brief of 22 February reports events through 28 February.

\begin{quote}
\small\itshape
(Feb.~22) Operation Epic Fury launches 28 February: US--Israeli strikes, Iran retaliates with missiles across the Gulf, and the Strait of Hormuz is effectively shut to commercial shipping.\\[2pt]
(Mar.~15) War enters a fourth week as a 48-hour Hormuz ultimatum is issued, Iran strikes southern Israel, and both sides escalate threats against energy and desalination infrastructure.\\[2pt]
(Apr.~5) A 7 April ceasefire is announced but Hormuz remains closed as more than 800 vessels stay stranded, and direct talks open in Islamabad.
\end{quote}

The value to a policymaker is not that the framework predicts the ceasefire or the ultimatum, but that within hours of each week's news it re-derives the full multi-model, multi-scenario reading on that information, and says plainly which quantities moved, why, and how confident the coverage behind them is.

\subsection{Under a substitute orchestrator}\label{sec:backends}

\begin{figure*}[t]
\centering
\begin{tikzpicture}
\begin{groupplot}[
    group style={group size=2 by 1, horizontal sep=1.9cm},
    width=0.52\textwidth, height=5.4cm,
    symbolic x coords={Sonnet-4.6,Llama-4-Scout,gpt-oss-120b,gpt-oss-20b},
    xtick=data, enlarge x limits=0.16,
    tick align=outside, tickpos=left,
    title style={font=\footnotesize\bfseries, anchor=south, yshift=1pt},
    ylabel style={font=\scriptsize}, xlabel style={font=\scriptsize},
    yticklabel style={font=\scriptsize},
    xticklabel style={font=\scriptsize, rotate=28, anchor=east},
    ymajorgrids, grid style={linegray!55, line width=0.3pt},
]
\nextgroupplot[
    title={(a) Mean wall-clock time per weekly re-run},
    ybar, bar width=13pt, ylabel={minutes}, ymin=0, ymax=30,
    nodes near coords, nodes near coords style={font=\tiny, /pgf/number format/fixed,
        /pgf/number format/precision=1},
    every node near coord/.append style={yshift=1pt},
]
\addplot[draw=none, fill=scB] coordinates {(Sonnet-4.6,26.3) (Llama-4-Scout,7.0) (gpt-oss-120b,8.5) (gpt-oss-20b,7.9)};
\nextgroupplot[
    title={(b) Model runs completed and run reliability},
    ybar, bar width=13pt, ylabel={avg.\ model runs completed}, ymin=0, ymax=48,
    legend style={font=\tiny, at={(0.5,1.24)}, anchor=south, legend columns=2,
                  draw=linegray, /tikz/every even column/.append style={column sep=5pt}},
]
\addplot[draw=none, fill=scC] coordinates {(Sonnet-4.6,18.0) (Llama-4-Scout,39.4) (gpt-oss-120b,36.0) (gpt-oss-20b,35.4)};
\legend{model runs completed}
\end{groupplot}
\begin{axis}[
    width=0.52\textwidth, height=5.4cm,
    at={(group c2r1.south east)}, anchor=south east,
    axis y line*=right, axis x line=none,
    symbolic x coords={Sonnet-4.6,Llama-4-Scout,gpt-oss-120b,gpt-oss-20b},
    enlarge x limits=0.16,
    ymin=0, ymax=1.05, ylabel={weekly success rate},
    ylabel style={font=\scriptsize}, yticklabel style={font=\scriptsize},
]
\addplot[amber, line width=0.9pt, mark=diamond*, mark size=2.4pt] coordinates {(Sonnet-4.6,1.0) (Llama-4-Scout,0.875) (gpt-oss-120b,0.75) (gpt-oss-20b,0.75)};
\node[anchor=north east, font=\tiny, align=right, inner sep=1.5pt] at (axis description cs:0.97,0.97)
  {\textcolor{amber}{$\blacklozenge$}~success rate (right)};
\end{axis}
\end{tikzpicture}
\caption{Substituting the orchestrator itself. Each of the four backends that ran executed the identical weekly temporal backfill over the demonstration window on the same per-week news feed. (a) Mean wall-clock time per weekly re-run: the two open models complete a full week in roughly seven to nine minutes against about twenty-six for \texttt{claude-sonnet-4.6}. (b) The mean number of (scenario, model) runs completed per weekly re-run (bars, left axis) against the fraction of weekly runs reaching a clean synthesis (markers, right axis). \texttt{claude-sonnet-4.6} never failed a week but completed fewer model runs on average, while the faster open backends completed roughly twice as many runs at a modestly lower weekly success rate. Four further backends produced no usable brief in this batch and are omitted; their failure was recorded and the batch continued rather than halting. All values are batch aggregates over the eight weekly runs.}
\label{fig:results_backends}
\end{figure*}
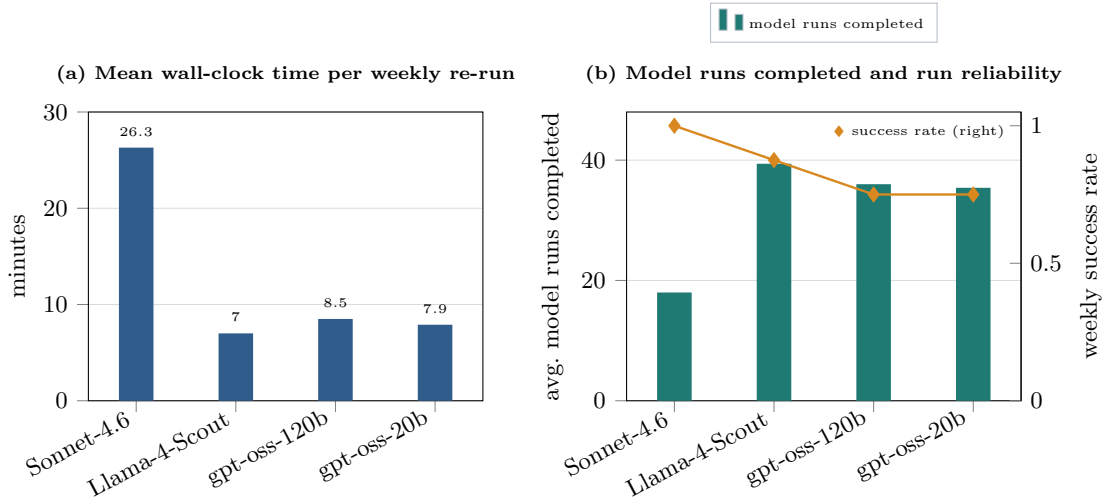

Because the orchestrating LLM is itself a configurable component, the same weekly window can be re-run under a different backend by changing a single setting. This is substitution applied to the orchestrator rather than to a domain model. We exercised it by running eight weekly re-runs under eight candidate backends drawn from two providers. Four produced complete analyses; four never returned a usable brief and were recorded as failed, and the batch continued without them. Because LLM capabilities are advancing rapidly, the orchestrating model should be as substitutable within the framework as the underlying economic models. Figure~\ref{fig:results_backends} summarizes the four that ran.

Two readings follow. First, the substitution is genuinely cheap: obtaining a second, independent orchestration of the entire weekly analysis required no code change, and the faster backends turn a full weekly re-analysis around in under ten minutes, well inside the tempo at which a crisis desk receives new information. Second, the backends trade off along an interpretable frontier rather than dominating one another. One reached a clean synthesis in every week but completed fewer model runs and ran slowest; the others ran three to four times faster and completed roughly twice as many model runs at a modestly lower weekly success rate. An analyst who needs a defensible reading every week and one who needs the broadest coverage at lowest latency are pointed toward different backends by the same comparison. All four that ran tracked the same qualitative crisis arc, so the choice is one of operating point, not of substantive conclusion.

\section{Discussion}\label{sec:discussion}

\subsection{Advantages of cheap substitution}\label{subsec:buys}

Crisis analysis has been constrained less by what models can compute than by what analysts can integrate under time constraints and policy pressure. We demonstrate the level of decision support that becomes available when the orchestration constraint is relaxed. Asking an analyst to run even a few energy models and a CGE quickly for a policymaker costs a weekend; running sixteen models across many commodities and five scenarios, with macroeconomic analysis, and refreshing everything weekly on new information would require a month of work. Within our framework the same task is routine: adding a model means writing one wrapper, and changing a scenario means editing one file. This flexibility lets analysts assemble pipelines that run ensembles to better characterize the range of plausible futures.

Breadth changes the range of outcomes decision-makers see. The non-monotone oil-price impulse is invisible one scenario at a time and becomes clear only when all five are compared. Cheap substitution also turns one-off studies into monitored trajectories, and it recasts divergence between models as a signal of differing assumptions rather than an anomaly to be averaged away. Because each figure carries its provenance, that signal is actionable: the analyst can follow an unexpected result back to the model and the assumption behind it, so the built-in limitations of any one model are offset by the others rather than propagating unseen into the answer. The same cheapness reaches the orchestrator itself: swapping the LLM backend for the entire weekly window is a one-line change (Section~\ref{sec:backends}). At a time of rising concern over LLM token costs, this flexibility also frees analysts from dependence on any one vendor.

\subsection{What the demonstration surfaced about the crisis}\label{subsec:surfaced}

The demonstration surfaced several readings worth recording, each stated as a portfolio output under a given scenario rather than a settled conclusion. The United States sits on the supply side of this disruption in a way it did not in the 1973 or 1979 shocks, and the CGE models put a sign and a rough magnitude on the resulting asymmetry: gains for Gulf producers against losses concentrated in import-dependent regions. The same direction recurs in fertilizer and helium, where US production and reserves leave those markets comparatively insulated. This differential between the US and the rest of the world appears in the modelling and has also been visible in markets since the strikes and closure.

The water models flag a potentially catastrophic risk to desalination capacity that conventional market analysis tended to underweight throughout the crisis. That this response lies dormant in the contained scenarios and activates only under escalation is by design, since the scenario-scoped forwarding rule routes the water signal into the macro layer only when narrative inputs stipulate infrastructure-level damage.

The macroeconomic readings are consistent with a contained rather than catastrophic US outcome under the matrix scenarios, deepening only under the prolonged-escalated and infrastructure-collapse tails. The structural reason for the mild impact on the United States comes from the unique position the United States holds as a belligerent in the conflict and a net exporter of the commodities most affected; the terms-of-trade gain may offset the inflationary drag. This reasoning remains contingent on the conflict staying contained and on the absence of a broader financial crisis, and the cross-scenario view is explicit that the tails tell a worse story. It should also be noted that long-term impacts from structural changes to the geopolitical landscape are not considered and may be substantial.

\subsection{Validity and the discipline of non-fabrication}\label{subsec:validity}

A framework that puts an LLM in the analytical loop invites the objection that fluency masks fabricated or silently harmonized numbers. The architecture forecloses this structurally rather than by exhortation: the LLM never originates a quantity, the forwarding mechanism overwrites narrative-derived values with model-derived ones and records the override, and the provenance chain makes every number traceable.

The clearest evidence that the discipline is more than nominal is the treatment of anomalies. The Scenario-A CGE outlier, the LNG-versus-gas spread, and the shipping disagreement were flagged and preserved, not smoothed into a plausible consensus. A system optimized for the appearance of coherence would have hidden them; a system optimized for validity does the opposite.

\subsection{Limitations}\label{subsec:limitations}

The limitations of the method are real, and the demonstration illustrates them sharply. Most models in this run operated as literature-calibrated reductions, so the numbers are order-of-magnitude readings whose role is to exercise the orchestration, not to forecast the crisis. The framework's value is that it makes such unreliability visible and cheap to correct by substituting a better model or improved parameterizations.

Regional resolution rests almost entirely on two CGE models, the scenario design embeds subjective judgements, and the strategic track is more qualitative than the rest. The agentic layer introduces abstraction in translating narratives into parameters; analyst checkpoints mitigate but do not eliminate the possibility of systematic extraction bias, which we have not formally characterized. The Lucas critique remains a biting challenge for ours and most similar micro-to-macro modelling exercises. Finally, the crisis is ongoing, and the analysis reflects information available at the time of writing.

\subsection{A portable contribution}\label{subsec:portable}

The broader contribution is portable and has two parts. First, linking models that already exist, rather than building from scratch, moves rigorous analysis close to the pace of the decision and replaces point forecasting with a broad view of possible outcomes, which is where unintended consequences become visible. Second, running a suite rather than a single model keeps the assumptions built into any one model from driving the result, surfaces divergence instead of averaging it, and lets the analyst trace a surprising output back to its source.

We develop an architecture in which models and scenarios are cheap to substitute and compose and analysis can be refreshed as information arrives. The architecture provides a template for rapid response to a wide class of energy and geopolitical disruptions. Outputs are verifiable and traceable through each level, with assumptions available to policymakers and models inheriting their pre-existing documentation.

The framework is also designed to improve over time. Each model can be promoted from a literature-calibrated reduction to a fully wired implementation, so instead of one-off assessments of individual events, every new model becomes an improvement the framework absorbs one adapter at a time.

\section{Methods}\label{sec:methods}

\subsection{Domain models and execution tiers}\label{subsec:methods_models}
The framework draws on a portfolio spanning the five commodity systems, shipping, macroeconomic general equilibrium, and long-run energy-systems optimization. Every model is wrapped by a typed adapter that declares a Pydantic input and output schema, resource requirements, and the analytical level at which it operates; adapters are registered in a central registry and dispatched by the orchestrator.

We distinguish four execution tiers. Fully wired adapters run the underlying model end-to-end with no extra configuration. Config-aware adapters run it when a configuration pointing at the required binary, dataset, or licence is supplied. Analytical-MVP adapters run a closed-form, literature-calibrated reduction that activates when no vendored binary is present, so each commodity system always returns a runnable output while the full upstream path is preserved. Typed stubs declare schemas pending upstream code.

At the time of writing the registry holds thirty-four adapters; the demonstration run drew output from sixteen, with the rest recorded as skipped. Failures are isolated: a model that cannot run is marked skipped and the pipeline continues, while the synthesis records which models contributed to each scenario. The oil layer uses the world-equilibrium model of Bornstein, Krusell and Rebelo \cite{bib_bornstein_oil}, POLES-JRC \cite{bib_poles}, MarketSim \cite{bib_marketsim}, and the Federal Reserve workhorse oil model \cite{bib_fed_oil}; the fertilizer and agricultural layer draws on CAPRI \cite{bib_capri}, MAgPIE \cite{bib_magpie}, GTAP \cite{bib_gtap}, and APSIM \cite{bib_apsim}; short-run macro is anchored by NEMS \cite{bib_nems}; and the water layer uses WEAP \cite{bib_weap}, WaterGAP2 \cite{bib_watergap}, CWatM, and SahysMod. Notably, some of the chosen models are not public: they are proprietary tools from working natural-gas analysts, delivered as spreadsheets. Our framework makes rapid integration of such models significantly more achievable. Appendix~\ref{secA1} gives the full inventory, platform, and execution status.

\subsection{Scenario construction}\label{subsec:methods_scenario}
We adopt a structured scenario methodology in the tradition of Schwartz \cite{bib_schwartz}, identifying predetermined elements and the two critical uncertainties that structure the matrix: closure duration and escalation scope. Crossing them yields the four matrix scenarios; a fifth, off-matrix tail-risk scenario stresses regional infrastructure destruction. For each scenario the LLM generates a narrative timeline, extracts quantitative parameters (disruption duration, production losses by commodity, price trajectories, rerouting costs, and infrastructure-damage assessments), and maps them to each model's input specification.

\subsection{Orchestration architecture}\label{subsec:methods_arch}
\begin{figure*}[t]
\centering
\resizebox{\textwidth}{!}{%
\begin{tikzpicture}[font=\sffamily,
    node distance=6mm and 8mm,
    stage/.style={draw=steel, line width=0.8pt, rounded corners=3pt, fill=paleblue,
      align=center, font=\sffamily\small, inner sep=5pt, minimum height=1.1cm, text width=2.35cm},
    llm/.style={stage, fill=paleteal, draw=teal},
    io/.style={stage, fill=palegray, draw=slate},
    ckpt/.style={draw=amber, line width=0.9pt, fill=paleamber, align=center,
      font=\sffamily\scriptsize\bfseries, shape=diamond, aspect=1.6, inner sep=1pt,
      text width=1.4cm, minimum height=1.1cm},
    support/.style={draw=slate, line width=0.7pt, dashed, rounded corners=3pt, fill=white,
      align=center, font=\sffamily\scriptsize},
    arr/.style={-{Stealth[length=2.6mm]}, line width=1pt, steel},
    supparr/.style={-{Stealth[length=2.2mm]}, line width=0.8pt, slate, dashed},
  ]
  \node[io]                   (crisis) {Crisis\\description};
  \node[llm, right=of crisis] (scen)   {Scenario\\generation\\\scriptsize(LLM)};
  \node[ckpt, right=of scen]  (c1)     {analyst\\\#1};
  \node[llm, right=of c1]     (param)  {Parameter\\extraction\\\scriptsize(LLM)};
  \node[ckpt, right=of param] (c2)     {analyst\\\#2};
  \node[stage, below=1.8cm of crisis] (exec) {Model executor\\\scriptsize level-by-level\\\scriptsize fan-out};
  \node[stage, right=of exec] (fwd)    {Upstream$\to$\\downstream\\forwarding};
  \node[stage, right=of fwd]  (cons)   {Cross-model\\consistency};
  \node[llm, right=of cons]   (synth)  {Synthesis\\\scriptsize(LLM,\\prose only)};
  \node[ckpt, right=of synth] (c3)     {analyst\\\#3};
  \node[io, right=of c3]      (report) {Report\\$+$ provenance};
  \draw[arr] (crisis) -- (scen);
  \draw[arr] (scen)   -- (c1);
  \draw[arr] (c1)     -- (param);
  \draw[arr] (param)  -- (c2);
  \draw[arr] (c2.south) -- ([yshift=9mm]exec.north -| c2) -| (exec.north);
  \draw[arr] (exec)  -- (fwd);
  \draw[arr] (fwd)   -- (cons);
  \draw[arr] (cons)  -- (synth);
  \draw[arr] (synth) -- (c3);
  \draw[arr] (c3)    -- (report);
  \node[support, text width=3.4cm, below=1.0cm of exec] (reg)
        {\textbf{Model registry}\\typed adapters\\\emph{swap a model $=$ register an adapter}};
  \draw[supparr] (reg.north) -- (exec.south);
  \node[support, text width=4.6cm, below=1.0cm of cons] (yaml)
        {\textbf{YAML configuration}\\parameters $\cdot$ tolerance bands $\cdot$ forwarding \& scenario rules\\\emph{swap a scenario / retune $=$ edit YAML}};
  \draw[supparr] (yaml.north) -- (cons.south);
  \draw[supparr] (yaml.north) to[out=150,in=-30] (fwd.south);
  \begin{scope}[on background layer]
    \node[draw=teal, line width=1pt, dashed, rounded corners=6pt,
          fit=(crisis)(c2)(exec)(report)(reg)(yaml), inner sep=11pt] (wrap) {};
  \end{scope}
  \draw[-{Stealth[length=2.8mm]}, line width=1pt, teal, dashed]
        (report.north) -- ([yshift=9mm]report.north |- wrap.north)
        -- node[midway, above=1pt, font=\sffamily\footnotesize\itshape, teal]
              {Temporal layer: weekly re-run on the updated information set}
           ([yshift=9mm]crisis.north |- wrap.north) -- (crisis.north);
\end{tikzpicture}}
\caption{The orchestration pipeline. A crisis description drives LLM scenario generation and parameter extraction (teal); the model executor then dispatches adapters level by level, forwards upstream outputs into downstream inputs, runs the cross-model consistency pass, and invokes LLM synthesis, which composes prose only. Three mandatory analyst checkpoints (amber diamonds) gate the workflow after scenario generation, after parameter extraction, and after synthesis. The two dashed boxes are the surfaces that make substitution inexpensive: the model registry (adding a model means registering one typed adapter) and the YAML configuration (re-specifying a scenario, a tolerance band, or a forwarding rule is a configuration edit, not a code change). The dashed teal wrapper denotes the temporal layer, which re-runs the entire pipeline on a weekly cadence as new information arrives.}
\label{fig:pipeline}
\end{figure*}
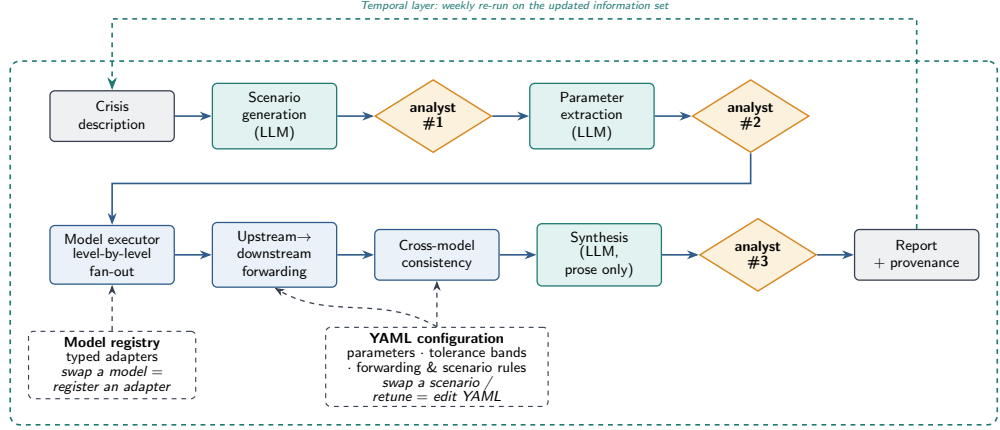

The pipeline is implemented on LangChain and its stateful-graph extension LangGraph \cite{bib_langchain}, which coordinates the workflow as a directed graph of typed states. Nodes correspond to scenario generation, parameter extraction, model execution, cross-model consistency, forwarding, and synthesis; per-field reducers let parallel branches accumulate results without overwriting. Algorithm~\ref{algo:pipeline} gives the canonical pseudocode and Fig.~\ref{fig:pipeline} renders the same pipeline as a process diagram.

Every domain model subclasses a model-adapter base class declaring validate, translate, execute, and parse methods, and six platform-specific base adapters (subprocess, GAMS, R, Julia, Excel, and AnyLogic) sit between it and each model so that a heterogeneous portfolio shares one interface. A single executor dispatches scenario-model pairs level by level, parallelizing within a level and enforcing sequential dependencies across levels.

Three analyst checkpoints are implemented as graph interrupts after scenario generation, after parameter extraction, and after synthesis; the pipeline has no autonomous mode. All configurable quantities live in YAML files, keeping the orchestration independent of any single LLM, solver, or deployment, and the same graph runs locally or on a SLURM cluster from one codebase.

\begin{algorithm}
\caption{Agentic AI orchestration pipeline}\label{algo:pipeline}
\begin{algorithmic}[1]
\Require Crisis description $\mathcal{C}$, domain models $\{M_1, \ldots, M_K\}$, scenario framework $\mathcal{F}$
\Ensure Synthesized multi-scenario, multi-scale analysis report $\mathcal{R}$
\State \textbf{Scenario generation:} $\{S_1, \ldots, S_N\} \leftarrow \text{LLM}(\mathcal{C}, \mathcal{F})$
\State \textbf{Analyst review:} $\{S_1, \ldots, S_N\} \leftarrow \text{Validate}(\{S_1, \ldots, S_N\})$
\For{each scenario $S_i$}
    \State \textbf{Parameter extraction:} $\theta_i \leftarrow \text{LLM}(S_i, \{M_1, \ldots, M_K\})$
    \State \textbf{Analyst validation:} $\theta_i \leftarrow \text{Validate}(\theta_i)$
    \For{each domain model $M_k$}
        \State \textbf{Input translation:} $x_{i,k} \leftarrow \text{Translate}(\theta_i, M_k)$
        \State \textbf{Model execution:} $y_{i,k} \leftarrow M_k(x_{i,k})$
    \EndFor
    \State \textbf{Cross-model consistency:} $\text{Flag}(\{y_{i,1}, \ldots, y_{i,K}\})$
    \State \textbf{Upstream-to-downstream forwarding:} overwrite downstream $\theta_i$ entries with upstream-derived values from completed $\{y_{i,k}\}$ per declarative mapping
    \State $y_{i,\text{macro}} \leftarrow M_{\text{CGE}}(\{y_{i,1}, \ldots, y_{i,K}\})$
\EndFor
\State \textbf{Synthesis:} $\mathcal{R} \leftarrow \text{LLM}(\{y_{i,k}\}_{i,k}, \{S_i\}_i, \theta)$
\State \textbf{Analyst review:} present $\mathcal{R}$ with full provenance
\end{algorithmic}
\end{algorithm}

\subsection{Connecting models across scales}\label{subsec:methods_connecting}
Quantitative linkages between levels are not re-derived by the LLM when a model has already computed them. A barrier-based forwarding mechanism fires at each level transition for which declarative mapping rules exist: combat outputs become physical disruption parameters for commodity models; the helium model's computed sectoral shortfall feeds the downstream semiconductor and helium-market models; and commodity price paths populate the macro models' multi-commodity shock vector.

Each rule specifies a downstream field, an ordered list of upstream sources with fallbacks, an optional transform, and an optional scenario restriction. The override policy is replace-with-metadata: an upstream value overwrites the LLM-extracted value, and the override is recorded with the original value, the source, and the symmetric percent deviation, which the consistency module flags when it exceeds a configurable threshold. When an upstream model is skipped, the LLM-extracted value is preserved. A scenario-scoped rule routes a water unmet-demand signal into the macro layer only under the infrastructure-collapse scenario.

\subsection{Non-fabrication and provenance}\label{subsec:methods_provenance}
The division of labor is enforced structurally. The LLM composes prose only; every quantitative value in the synthesis is passed through from a model run without modification, interpolation, or rounding. A provenance tracker records, for each result, the chain from synthesis output back through model output, model input, extracted parameter, and source narrative. A cross-model consistency pass reads a declarative tolerance-band rules file and raises explicit flags where overlapping outputs diverge, so disagreements appear in the synthesis rather than being harmonized. Every flagged disagreement, skipped adapter, override, and analyst correction is written to the provenance record.

\subsection{Temporal layer}\label{subsec:methods_temporal}
The temporal layer wraps the pipeline without modifying it, re-running it weekly on an updated information set. Each week, articles from commercial news interfaces and EIA indicator series (WTI, Brent, US crude stocks, and Henry Hub) are summarized by an LLM into a structured weekly brief conditioned on the previous week's brief. The summarizer may arrange source-attested quantities but may not originate numbers; any value in a brief traces to a specific article or series. The brief enters the pipeline through an augmented crisis description; scenario identifiers, registry, forwarding rules, and tolerances are unchanged, so weekly runs are directly comparable. All three checkpoints remain active each week, and an optional fourth review of the brief can precede a full re-run.

\backmatter

\bmhead{Supplementary information}
Supplementary materials include detailed scenario narratives, full parameter-extraction tables, individual model configuration files, and extended results tables for all domain models and scenarios.

\bmhead{Acknowledgements} The views and conclusions contained herein are those of the authors and should not be interpreted as necessarily representing the official policies or endorsements, either expressed or implied, of the United States Government (USG). Any opinions, findings and conclusions or recommendations expressed in this material are those of the author(s) and do not necessarily reflect the views of the National Science Foundation.

\section*{Declarations}

\begin{itemize}
\item \textbf{Funding:} This work was partially supported by the National Science Foundation under Grant No. NRT-HDR 2125295, the U.S. Department of Agriculture, National Institute of Food and
Agriculture (USDA NIFA) under award number 2023-68012-39076, and funding from a U.S. Government agency. The funding agencies did not play any role in the design of the study, analysis, interpretation of results, or the preparation of the manuscript.
\item \textbf{Competing interests:} The authors declare no competing interests.
\item \textbf{Ethics approval:} Not applicable.
\item \textbf{Data availability:} The indicator series used by the temporal layer are available from the US Energy Information Administration Open Data interface. Scenario narratives, extracted-parameter tables, and per-model configuration files are available upon request.
\item \textbf{Code availability:} The agentic orchestration framework and model-interface code will be made available at \href{https://github.com/jgolden36/agenticAIForScenarioAnalysis}{Agentic AI for Scenario Analysis Repository}. Individual domain models are subject to their respective licensing terms.
\item \textbf{Author contributions:} DG: Conceptualization, Formal Analysis, Methodology, Software, Visualization, Writing -- original draft, Writing -- review \& editing. BI: Conceptualization, Writing -- review \& editing. LV: Conceptualization, Funding Acquisition, Supervision, Validation, Writing -- original draft, Writing -- review \& editing. CDM: Conceptualization, Writing -- review \& editing. ST: Conceptualization, Funding Acquisition, Supervision, Validation, Writing -- original draft, Writing -- review \& editing.
\end{itemize}

\begin{appendices}

\section{Domain model inventory}\label{secA1}

Table~\ref{tab:model_inventory} lists every registered adapter by commodity system, analytical platform, and execution status. \textbf{Real}: the adapter runs the underlying model end-to-end in any valid configuration. \textbf{Real (cfg)}: it runs the underlying model when a YAML configuration pointing at the required binary, dataset, or licence is supplied. \textbf{Real (MVP)}: it runs a closed-form, literature-calibrated reduction that activates when the vendored binary is not available; the upstream path is preserved and activates when the assets are supplied. \textbf{Cfg stub}: dispatch machinery is wired and a YAML-activated execute path is in place but awaits upstream code or licence. \textbf{Stub}: a typed adapter with input and output schemas only.

\begin{longtable}{>{\raggedright\arraybackslash}p{2.2cm} >{\raggedright\arraybackslash}p{3.7cm} >{\raggedright\arraybackslash}p{2.6cm} >{\raggedright\arraybackslash}p{1.7cm} >{\raggedright\arraybackslash}p{2.9cm}}
\caption{Domain model inventory by commodity system, analytical platform, and current implementation status.}\label{tab:model_inventory} \\
\toprule
\thead{Domain} & \thead{Model} & \thead{Type / \\ Platform} & \thead{Status} & \thead{Role in framework} \\
\midrule
\endfirsthead
\multicolumn{5}{l}{\emph{Table~\ref{tab:model_inventory} continued}} \\
\toprule
\thead{Domain} & \thead{Model} & \thead{Type / \\ Platform} & \thead{Status} & \thead{Role in framework} \\
\midrule
\endhead
\midrule
\multicolumn{5}{r}{\emph{Continued on next page}} \\
\endfoot
\bottomrule
\endlastfoot

\multirow{4}{2.2cm}{Water}
  & WEAP / WEAP--MENA \cite{bib_weap} & Simulation; Windows COM / analytical fallback & Real (MVP) & Integrated water-resource planning; Gulf configuration \\
  & SahysMod & Simulation; native CLI & Real (cfg) & Spatially distributed agro-hydro-salinity modelling \\
  & WaterGAP2 \cite{bib_watergap} & Gridded global; binary or HTTP API & Cfg stub & Global gridded hydrological modelling of disruption \\
  & CWatM & Gridded global; Python / analytical fallback & Real (cfg) + Real (MVP) & Community-scale water availability under disruption \\
\midrule

\multirow{4}{2.2cm}{Oil}
  & World Equilibrium Oil Model (BKR) \cite{bib_bornstein_oil} & Structural GE; Octave + Dynare & Real (cfg) & Supply-disruption analysis in general equilibrium \\
  & POLES-JRC \cite{bib_poles} & Partial eq.; closed-form fallback & Real (MVP) & Global energy supply and demand; elasticity price impulse \\
  & MarketSim \cite{bib_marketsim} & Partial eq.; Python / Excel-VBA (upstream) & Real (MVP) & Consumer surplus and energy substitution \\
  & Fed Workhorse Oil Model \cite{bib_fed_oil} & Macro--energy; Python / MATLAB-R (upstream) & Real (MVP) & US monetary transmission of oil-price shocks \\
\midrule

\multirow{4}{2.2cm}{LNG}
  & Energy Flux Gas-Power Build-Out & Proprietary calc; Python & Real & US gas-to-power capacity constraints \\
  & Energy Flux LNG War-Profits & Proprietary calc; Python & Real & LNG export revenue under conflict scenarios \\
  & Global Gas Model (GGM) & Optimization; GAMS + CPLEX & Real (cfg) & Global gas trade-flow optimization \\
  & LNG Spreadsheet Tool (LNGST) & Spreadsheet; Python / Excel & Real (MVP) & Scenario-level LNG trade-flow simulation \\
\midrule

\multirow{3}{2.2cm}{Helium \& Semiconductors}
  & World Helium Model & Market eq.; Qatar-share closed form & Real (MVP) & Global helium supply--demand equilibrium \\
  & Argonne Helium ABM \cite{bib_argonne_helium} & Agent-based; Python / AnyLogic (JAR) & Real (MVP) & Contemporary helium-market dynamics \\
  & SimRLFab & RL / SimPy; Python 3.6 venv & Real (cfg) & Semiconductor fabrication disruption impacts \\
\midrule

\multirow{7}{2.2cm}{Fertilizer \& Agricultural Trade}
  & CAPRI \cite{bib_capri} & Partial eq.; GAMS & Stub & Regional agricultural policy impact \\
  & MAgPIE \cite{bib_magpie} & Optimization; R + GAMS & Real (cfg) & Land-use and agricultural production \\
  & SIMPLE-G & CGE; platform TBD & Stub & General-equilibrium agricultural trade \\
  & World Fertilizer Model & Market eq.; Python / GAMS (upstream) & Real (MVP) & Global fertilizer supply--demand \\
  & GTAP \cite{bib_gtap} & CGE; GEMPACK & Stub & Global agricultural and commodity trade \\
  & APSIM \cite{bib_apsim} & Crop simulation; Python / native CLI (upstream) & Real (MVP) & Physical crop-yield response \\
  & Futures forecasting (plotted as ``Fwd-curve'') & Time series; Python AR(1) & Real (MVP) & Commodity futures price trajectories \\
\midrule

\multirow{2}{2.2cm}{Shipping}
  & AISdb & Spatial DB; Python & Real (MVP) & AIS vessel tracking and rerouting calibration \\
  & aisstream (plotted as ``AIS-alt'') & Spatial analysis; Python + WebSocket & Real (MVP) & Transit time and fleet utilization \\
\midrule

\multirow{7}{2.2cm}{Macroeconomic / General Equilibrium}
  & NEMS \cite{bib_nems} & Systems model; Fortran + GAMS + Python & Real & National energy--economy projections \\
  & MAM & Macro econometric; EViews & Real & US macroeconomic feedback from energy prices \\
  & NREL baseline & Sectoral; Python (analytical MVP) & Real (MVP) & Electricity-sector baseline and disruption \\
  & MPSGE.jl & CGE; Python / Julia & Real (MVP) & General-equilibrium trade and welfare \\
  & OpenCGE (OG-Core / OG-USA) & CGE; Python + Dask & Real & Open-source dynamic OLG CGE \\
  & pycge & CGE; Python & Real (MVP) & SAM-driven CGE for sensitivity analysis \\
  & MIRAGRODEP & CGE; GAMS & Real & Multi-region CGE with agricultural--trade linkages \\
\midrule

\multirow{3}{2.2cm}{Energy Systems (long-run)}
  & OSeMOSYS & LP; GNU MathProg + GLPK/CBC & Real (cfg) & Open-source long-run energy-systems optimization \\
  & MESSAGEix & MIP; IIASA message-ix & Real (cfg) & Integrated assessment / energy-systems framework \\
  & TEMOA & LP; Pyomo + CBC & Real (cfg) & Tools for energy model optimization and analysis \\

\end{longtable}

\section{Coverage map}\label{secA4}

Figure~\ref{fig:results_coverage} records which of the sixteen models produced output under each scenario. Coverage was richest where the scenario admitted the most relevant models and thinnest at the matrix corners.

\begin{figure*}[t]
\centering
\begin{tikzpicture}[x=0.62cm, y=0.62cm]
\fill[covfill, draw=white, line width=1.1pt] (0,4) rectangle (1,5);
\node[font=\scriptsize, white] at (0.5,4.5) {\checkmark};
\fill[covfill, draw=white, line width=1.1pt] (0,3) rectangle (1,4);
\node[font=\scriptsize, white] at (0.5,3.5) {\checkmark};
\fill[covfill, draw=white, line width=1.1pt] (0,2) rectangle (1,3);
\node[font=\scriptsize, white] at (0.5,2.5) {\checkmark};
\fill[covfill, draw=white, line width=1.1pt] (0,1) rectangle (1,2);
\node[font=\scriptsize, white] at (0.5,1.5) {\checkmark};
\fill[covfill, draw=white, line width=1.1pt] (0,0) rectangle (1,1);
\node[font=\scriptsize, white] at (0.5,0.5) {\checkmark};
\fill[covfill, draw=white, line width=1.1pt] (1,4) rectangle (2,5);
\node[font=\scriptsize, white] at (1.5,4.5) {\checkmark};
\fill[covfill, draw=white, line width=1.1pt] (1,3) rectangle (2,4);
\node[font=\scriptsize, white] at (1.5,3.5) {\checkmark};
\fill[covfill, draw=white, line width=1.1pt] (1,2) rectangle (2,3);
\node[font=\scriptsize, white] at (1.5,2.5) {\checkmark};
\fill[covfill, draw=white, line width=1.1pt] (1,1) rectangle (2,2);
\node[font=\scriptsize, white] at (1.5,1.5) {\checkmark};
\fill[covfill, draw=white, line width=1.1pt] (1,0) rectangle (2,1);
\node[font=\scriptsize, white] at (1.5,0.5) {\checkmark};
\fill[covfill, draw=white, line width=1.1pt] (2,4) rectangle (3,5);
\node[font=\scriptsize, white] at (2.5,4.5) {\checkmark};
\fill[covfill, draw=white, line width=1.1pt] (2,3) rectangle (3,4);
\node[font=\scriptsize, white] at (2.5,3.5) {\checkmark};
\fill[covfill, draw=white, line width=1.1pt] (2,2) rectangle (3,3);
\node[font=\scriptsize, white] at (2.5,2.5) {\checkmark};
\fill[covfill, draw=white, line width=1.1pt] (2,1) rectangle (3,2);
\node[font=\scriptsize, white] at (2.5,1.5) {\checkmark};
\fill[covfill, draw=white, line width=1.1pt] (2,0) rectangle (3,1);
\node[font=\scriptsize, white] at (2.5,0.5) {\checkmark};
\fill[covfill, draw=white, line width=1.1pt] (3,4) rectangle (4,5);
\node[font=\scriptsize, white] at (3.5,4.5) {\checkmark};
\fill[covfill, draw=white, line width=1.1pt] (3,3) rectangle (4,4);
\node[font=\scriptsize, white] at (3.5,3.5) {\checkmark};
\fill[covfill, draw=white, line width=1.1pt] (3,2) rectangle (4,3);
\node[font=\scriptsize, white] at (3.5,2.5) {\checkmark};
\fill[covfill, draw=white, line width=1.1pt] (3,1) rectangle (4,2);
\node[font=\scriptsize, white] at (3.5,1.5) {\checkmark};
\fill[covfill, draw=white, line width=1.1pt] (3,0) rectangle (4,1);
\node[font=\scriptsize, white] at (3.5,0.5) {\checkmark};
\fill[covfill, draw=white, line width=1.1pt] (4,4) rectangle (5,5);
\node[font=\scriptsize, white] at (4.5,4.5) {\checkmark};
\fill[covfill, draw=white, line width=1.1pt] (4,3) rectangle (5,4);
\node[font=\scriptsize, white] at (4.5,3.5) {\checkmark};
\fill[covfill, draw=white, line width=1.1pt] (4,2) rectangle (5,3);
\node[font=\scriptsize, white] at (4.5,2.5) {\checkmark};
\fill[covfill, draw=white, line width=1.1pt] (4,1) rectangle (5,2);
\node[font=\scriptsize, white] at (4.5,1.5) {\checkmark};
\fill[covfill, draw=white, line width=1.1pt] (4,0) rectangle (5,1);
\node[font=\scriptsize, white] at (4.5,0.5) {\checkmark};
\fill[covempty, draw=white, line width=1.1pt] (5,4) rectangle (6,5);
\fill[covfill, draw=white, line width=1.1pt] (5,3) rectangle (6,4);
\node[font=\scriptsize, white] at (5.5,3.5) {\checkmark};
\fill[covfill, draw=white, line width=1.1pt] (5,2) rectangle (6,3);
\node[font=\scriptsize, white] at (5.5,2.5) {\checkmark};
\fill[covempty, draw=white, line width=1.1pt] (5,1) rectangle (6,2);
\fill[covempty, draw=white, line width=1.1pt] (5,0) rectangle (6,1);
\fill[covempty, draw=white, line width=1.1pt] (6,4) rectangle (7,5);
\fill[covfill, draw=white, line width=1.1pt] (6,3) rectangle (7,4);
\node[font=\scriptsize, white] at (6.5,3.5) {\checkmark};
\fill[covfill, draw=white, line width=1.1pt] (6,2) rectangle (7,3);
\node[font=\scriptsize, white] at (6.5,2.5) {\checkmark};
\fill[covempty, draw=white, line width=1.1pt] (6,1) rectangle (7,2);
\fill[covempty, draw=white, line width=1.1pt] (6,0) rectangle (7,1);
\fill[covfill, draw=white, line width=1.1pt] (7,4) rectangle (8,5);
\node[font=\scriptsize, white] at (7.5,4.5) {\checkmark};
\fill[covempty, draw=white, line width=1.1pt] (7,3) rectangle (8,4);
\fill[covfill, draw=white, line width=1.1pt] (7,2) rectangle (8,3);
\node[font=\scriptsize, white] at (7.5,2.5) {\checkmark};
\fill[covempty, draw=white, line width=1.1pt] (7,1) rectangle (8,2);
\fill[covempty, draw=white, line width=1.1pt] (7,0) rectangle (8,1);
\fill[covempty, draw=white, line width=1.1pt] (8,4) rectangle (9,5);
\fill[covfill, draw=white, line width=1.1pt] (8,3) rectangle (9,4);
\node[font=\scriptsize, white] at (8.5,3.5) {\checkmark};
\fill[covfill, draw=white, line width=1.1pt] (8,2) rectangle (9,3);
\node[font=\scriptsize, white] at (8.5,2.5) {\checkmark};
\fill[covempty, draw=white, line width=1.1pt] (8,1) rectangle (9,2);
\fill[covempty, draw=white, line width=1.1pt] (8,0) rectangle (9,1);
\fill[covempty, draw=white, line width=1.1pt] (9,4) rectangle (10,5);
\fill[covfill, draw=white, line width=1.1pt] (9,3) rectangle (10,4);
\node[font=\scriptsize, white] at (9.5,3.5) {\checkmark};
\fill[covfill, draw=white, line width=1.1pt] (9,2) rectangle (10,3);
\node[font=\scriptsize, white] at (9.5,2.5) {\checkmark};
\fill[covempty, draw=white, line width=1.1pt] (9,1) rectangle (10,2);
\fill[covempty, draw=white, line width=1.1pt] (9,0) rectangle (10,1);
\fill[covfill, draw=white, line width=1.1pt] (10,4) rectangle (11,5);
\node[font=\scriptsize, white] at (10.5,4.5) {\checkmark};
\fill[covempty, draw=white, line width=1.1pt] (10,3) rectangle (11,4);
\fill[covfill, draw=white, line width=1.1pt] (10,2) rectangle (11,3);
\node[font=\scriptsize, white] at (10.5,2.5) {\checkmark};
\fill[covempty, draw=white, line width=1.1pt] (10,1) rectangle (11,2);
\fill[covempty, draw=white, line width=1.1pt] (10,0) rectangle (11,1);
\fill[covfill, draw=white, line width=1.1pt] (11,4) rectangle (12,5);
\node[font=\scriptsize, white] at (11.5,4.5) {\checkmark};
\fill[covempty, draw=white, line width=1.1pt] (11,3) rectangle (12,4);
\fill[covfill, draw=white, line width=1.1pt] (11,2) rectangle (12,3);
\node[font=\scriptsize, white] at (11.5,2.5) {\checkmark};
\fill[covempty, draw=white, line width=1.1pt] (11,1) rectangle (12,2);
\fill[covempty, draw=white, line width=1.1pt] (11,0) rectangle (12,1);
\fill[covempty, draw=white, line width=1.1pt] (12,4) rectangle (13,5);
\fill[covfill, draw=white, line width=1.1pt] (12,3) rectangle (13,4);
\node[font=\scriptsize, white] at (12.5,3.5) {\checkmark};
\fill[covfill, draw=white, line width=1.1pt] (12,2) rectangle (13,3);
\node[font=\scriptsize, white] at (12.5,2.5) {\checkmark};
\fill[covempty, draw=white, line width=1.1pt] (12,1) rectangle (13,2);
\fill[covempty, draw=white, line width=1.1pt] (12,0) rectangle (13,1);
\fill[covempty, draw=white, line width=1.1pt] (13,4) rectangle (14,5);
\fill[covfill, draw=white, line width=1.1pt] (13,3) rectangle (14,4);
\node[font=\scriptsize, white] at (13.5,3.5) {\checkmark};
\fill[covfill, draw=white, line width=1.1pt] (13,2) rectangle (14,3);
\node[font=\scriptsize, white] at (13.5,2.5) {\checkmark};
\fill[covempty, draw=white, line width=1.1pt] (13,1) rectangle (14,2);
\fill[covempty, draw=white, line width=1.1pt] (13,0) rectangle (14,1);
\fill[covempty, draw=white, line width=1.1pt] (14,4) rectangle (15,5);
\fill[covempty, draw=white, line width=1.1pt] (14,3) rectangle (15,4);
\fill[covfill, draw=white, line width=1.1pt] (14,2) rectangle (15,3);
\node[font=\scriptsize, white] at (14.5,2.5) {\checkmark};
\fill[covempty, draw=white, line width=1.1pt] (14,1) rectangle (15,2);
\fill[covempty, draw=white, line width=1.1pt] (14,0) rectangle (15,1);
\fill[covempty, draw=white, line width=1.1pt] (15,4) rectangle (16,5);
\fill[covempty, draw=white, line width=1.1pt] (15,3) rectangle (16,4);
\fill[covfill, draw=white, line width=1.1pt] (15,2) rectangle (16,3);
\node[font=\scriptsize, white] at (15.5,2.5) {\checkmark};
\fill[covempty, draw=white, line width=1.1pt] (15,1) rectangle (16,2);
\fill[covempty, draw=white, line width=1.1pt] (15,0) rectangle (16,1);
\node[anchor=east, font=\footnotesize\bfseries] at (-0.15,4.5) {A};
\node[anchor=east, font=\footnotesize\bfseries] at (-0.15,3.5) {B};
\node[anchor=east, font=\footnotesize\bfseries] at (-0.15,2.5) {C};
\node[anchor=east, font=\footnotesize\bfseries] at (-0.15,1.5) {D};
\node[anchor=east, font=\footnotesize\bfseries] at (-0.15,0.5) {E};
\node[anchor=west, font=\scriptsize] at (16.15,4.5) {8};
\node[anchor=west, font=\scriptsize] at (16.15,3.5) {11};
\node[anchor=west, font=\scriptsize] at (16.15,2.5) {16};
\node[anchor=west, font=\scriptsize] at (16.15,1.5) {5};
\node[anchor=west, font=\scriptsize] at (16.15,0.5) {5};
\node[anchor=south, font=\tiny\itshape, slate, align=center] at (16.55,5.05) {models};
\node[rotate=90, anchor=east, font=\scriptsize\bfseries] at (0.5,-0.12) {Fed oil};
\node[rotate=90, anchor=east, font=\scriptsize\bfseries] at (1.5,-0.12) {EIA macro (MAM)};
\node[rotate=90, anchor=east, font=\scriptsize\bfseries] at (2.5,-0.12) {MPSGE.jl};
\node[rotate=90, anchor=east, font=\scriptsize\bfseries] at (3.5,-0.12) {NREL elec.};
\node[rotate=90, anchor=east, font=\scriptsize\bfseries] at (4.5,-0.12) {pycge};
\node[rotate=90, anchor=east, font=\scriptsize] at (5.5,-0.12) {AIS-alt};
\node[rotate=90, anchor=east, font=\scriptsize] at (6.5,-0.12) {AISdb};
\node[rotate=90, anchor=east, font=\scriptsize] at (7.5,-0.12) {Energy~Flux};
\node[rotate=90, anchor=east, font=\scriptsize] at (8.5,-0.12) {Fwd-curve};
\node[rotate=90, anchor=east, font=\scriptsize] at (9.5,-0.12) {LNGST};
\node[rotate=90, anchor=east, font=\scriptsize] at (10.5,-0.12) {MarketSim};
\node[rotate=90, anchor=east, font=\scriptsize] at (11.5,-0.12) {POLES-JRC};
\node[rotate=90, anchor=east, font=\scriptsize] at (12.5,-0.12) {World Fert.};
\node[rotate=90, anchor=east, font=\scriptsize] at (13.5,-0.12) {World Helium};
\node[rotate=90, anchor=east, font=\scriptsize] at (14.5,-0.12) {CWatM};
\node[rotate=90, anchor=east, font=\scriptsize] at (15.5,-0.12) {WEAP--MENA};
\draw[decorate, decoration={brace, amplitude=4pt, mirror}, slate, line width=0.6pt] (0.05,5.12) -- (4.95,5.12);
\node[anchor=south, font=\tiny\itshape, slate] at (2.5,5.42) {all-scenario anchor models};
\fill[covfill] (3.0,-4.9) rectangle (3.5,-4.35); \node[anchor=west, font=\scriptsize] at (3.65,-4.62) {produced output};
\fill[covempty] (8.4,-4.9) rectangle (8.9,-4.35); \node[anchor=west, font=\scriptsize] at (9.05,-4.62) {not dispatched};
\end{tikzpicture}
\caption{Coverage and graceful degradation. The grid records which of the sixteen models produced output under each scenario (filled) and which were not dispatched (empty). Per-scenario totals on the right (8, 11, 16, 5, 5) give the number of contributing models; the five all-scenario anchor models (braced), namely the Federal Reserve oil model, the EIA macroeconomic activity module, MPSGE.jl, the NREL electricity baseline, and pycge, are present under every scenario and so underpin every cross-scenario comparison. Coverage is richest under Scenario~C and thinnest under Scenarios~D and~E.}
\label{fig:results_coverage}
\end{figure*}

\end{appendices}


\begin{thebibliography}{99}

\bibitem{bib_hourcade2006}
Hourcade, J.-C., Jaccard, M., Bataille, C. \& Ghersi, F.
Hybrid modeling: new answers to old challenges. \textit{Energy J.} \textbf{27}, 1--11 (2006).

\bibitem{bib_eia_hormuz}
U.S. Energy Information Administration.
The Strait of Hormuz is the world's most important oil transit chokepoint.
\textit{Today in Energy} (2023).

\bibitem{bib_eia_q1_2026}
U.S. Energy Information Administration.
\newblock Crude oil and petroleum product prices increased sharply in the first quarter of 2026.
\newblock \emph{Today in Energy}, 7 April 2026.
\newblock \url{https://www.eia.gov/todayinenergy/detail.php?id=67424} (accessed 13 July 2026).

\bibitem{bib_hormuz_strikes}
Maritime Reporter / MarineLink.
\newblock The choking point: how Strait of Hormuz disruptions are reshaping global shipping.
\newblock 28 May 2026.
\newblock \url{https://www.marinelink.com/news/choking-point-strait-hormuz-disruptions-539695} (accessed 13 July 2026).

\bibitem{bib_reuters_shutins}
Reuters.
\newblock Saudi Arabia cuts oil output to 8 million bpd amid Iran war, sources say.
\newblock 13 March 2026.
\newblock \url{https://www.reuters.com/business/energy/saudi-arabia-cuts-oil-output-20-8-million-bpd-amid-iran-war-sources-say-2026-03-13/} (accessed 13 July 2026).

\bibitem{bib_brent_peak}
Discovery Alert.
\newblock Brent oil price surge: Strait of Hormuz blockade crisis 2026.
\newblock 1 May 2026.
\newblock \url{https://discoveryalert.com.au/brent-oil-price-surge-strait-hormuz-blockade-2026/} (accessed 13 July 2026).

\bibitem{bib_ttf_gas}
International Road Transport Union (IRU).
\newblock More turmoil: early pump price movements seen globally.
\newblock 26 March 2026.
\newblock \url{https://www.iru.org/news-resources/newsroom/more-turmoil-early-pump-price-movements-seen-globally} (accessed 13 July 2026).

\bibitem{bib_warrisk}
Sea Vantage.
\newblock Strait of Hormuz crisis 2026: full timeline and ocean freight impact.
\newblock 12 April 2026.
\newblock \url{https://www.seavantage.com/blog/strait-of-hormuz-crisis-2026-shipping-disruption-timeline} (accessed 13 July 2026).

\bibitem{bib_schwartz}
Schwartz, P.
\textit{The Art of the Long View: Planning for the Future in an Uncertain World} (Currency Doubleday, 1991).

\bibitem{bib_dod_ddd}
Singer, P.~W.
\textit{Wired for War: The Robotics Revolution and Conflict in the 21st Century} (Penguin, 2009).

\bibitem{bib_bornstein_oil}
Bornstein, G., Krusell, P. \& Rebelo, S.
A world equilibrium model of the oil market.
\textit{Rev. Econ. Stud.} \textbf{90}, 822--852 (2023).

\bibitem{bib_fed_oil}
Baumeister, C. \& Hamilton, J.~D.
Structural interpretation of vector autoregressions with incomplete identification: revisiting the role of oil supply and demand shocks.
\textit{Am. Econ. Rev.} \textbf{109}, 1873--1910 (2019).

\bibitem{bib_marketsim}
U.S. Bureau of Ocean Energy Management.
Consumer surplus and energy substitutes for OCS oil and gas production: the 2021 revised Market Simulation Model (MarketSim).
\textit{BOEM Technical Report} (2021).

\bibitem{bib_poles}
Criqui, P. \& Mima, S.
European climate--energy security nexus: a model based scenario analysis.
\textit{Energy Policy} \textbf{41}, 827--842 (2012).

\bibitem{bib_argonne_helium}
Hamlin, T. \& Paoli, R.
Development of an agent-based model to analyze contemporary helium markets.
\textit{Argonne National Laboratory Technical Report} ANL-22/XX (2022).

\bibitem{bib_weap}
Sieber, J. \& Purkey, D.
\textit{Water Evaluation and Planning System (WEAP): User Guide} (Stockholm Environment Institute, 2015).

\bibitem{bib_watergap}
M{\"u}ller Schmied, H. et al.
Variations of global and continental water balance components as impacted by climate forcing uncertainty and human water use.
\textit{Hydrol. Earth Syst. Sci.} \textbf{18}, 3351--3366 (2014).

\bibitem{bib_capri}
Britz, W. \& Witzke, P.
CAPRI model documentation.
\textit{University of Bonn Technical Report} (2014).

\bibitem{bib_magpie}
Dietrich, J.~P. et al.
MAgPIE 4: a modular open-source framework for modeling global land systems.
\textit{Geosci. Model Dev.} \textbf{12}, 1299--1317 (2019).

\bibitem{bib_gtap}
Hertel, T.~W. (ed.)
\textit{Global Trade Analysis: Modeling and Applications} (Cambridge University Press, 1997).

\bibitem{bib_apsim}
Holzworth, D.~P. et al.
APSIM: evolution towards a new generation of agricultural systems simulation.
\textit{Environ. Model. Softw.} \textbf{62}, 327--350 (2014).

\bibitem{bib_nems}
U.S. Energy Information Administration.
The National Energy Modeling System: an overview.
\textit{EIA Technical Report} DOE/EIA-0581 (2023).

\bibitem{bib_langchain}
Chase, H. et al.
LangChain: building applications with LLMs through composability.
\textit{GitHub} \url{https://github.com/langchain-ai/langchain} (2024).

\end{thebibliography}
\end{document}